\documentstyle[myart,amsfonts,12pt]{article}

\normalbaselines
\font\tenbm=cmmib10
\font\sevenbm=cmmib7
\font\fivebm=cmmib5
\newfam\bmfam
\textfont\bmfam=\tenbm \scriptfont\bmfam=\sevenbm
\scriptscriptfont\bmfam=\fivebm
{\count0=\number\bmfam \multiply\count0 by "100
\def\defbgreek#1#2#3{{\count1=\count0 \advance\count1 by "#2#3
  \global\mathchardef#1=\count1 }}
\defbgreek\balpha  0B \defbgreek\brho       1A
\defbgreek\bbeta   0C \defbgreek\bsigma     1B
\defbgreek\bgamma  0D \defbgreek\btau       1C
\defbgreek\bdelta  0E \defbgreek\bupsilon   1D
\defbgreek\bepsilon0F \defbgreek\bphi       1E
\defbgreek\bzeta   10 \defbgreek\bchi       1F
\defbgreek\bmeta   11 \defbgreek\bpsi       20
\defbgreek\btheta  12 \defbgreek\bomega     21
\defbgreek\biota   13 \defbgreek\bvarepsilon22
\defbgreek\bkappa  14 \defbgreek\bvartheta  23
\defbgreek\blambda 15 \defbgreek\bvarpi     24
\defbgreek\bmu     16 \defbgreek\bvarrho    25
\defbgreek\bnu     17 \defbgreek\bvarsigma  26
        \defbgreek\bxi     18 \defbgreek\bvarphi    27
\defbgreek\bpi     19}
\input tcilatex

\begin{document}

\author{Yuri. A. Rylov}
\title{Ptolemyness of conventional program for microcosm investigations and
alternative research program}
\date{Institute for Problems in Mechanics, Russian Academy of Sciences, \\
101-1 Vernadskii Ave., Moscow, 117526, Russia \\
email: rylov@ipmnet.ru}
\maketitle

\begin{abstract}
It is shown that the conventional approach to microcosm investigations uses
an incorrect supposition (incorrect space-time model) whose incorrectness is
compensated by means of additional hypotheses (quantum mechanics
principles). Such a conception reminds the Ptolemaic doctrine of celestial
mechanics. Alternative research program, which uses more correct space-time
model and does not need additional hypotheses (quantum mechanics principles)
for free explanation of quantum effects, is suggested. If the more correct
space-time model were known in the beginning of XXth century, when research
of microcosm started, quantum mechanics could develop in other way. The
alternative research program appeared with a secular delay, because all this
time the necessary mathematical technique was not available for researchers.
Absence of necessary mathematical technique is connected with some
prejudices which have been overcame at the construction of new conception of
geometry and that of statistical description. Basic statements of the new
mathematical technique and principles of its application in alternative
research program are presented in the paper.
\end{abstract}


\newpage

\section{Introduction}

When in the beginning of XX century one starts to investigate physical
phenomena in microcosm, researchers met two serious problems, which could
not be solved in the scope of the classical physics of that time. They
demanded a new approaches. The first problem is the problem of microparticle
motion with velocities close to the speed of the light. This problem had
been solved by construction of special relativity theory. The concentrated
expression of the relativity principles is the statement that the event
space (space-time) is described by the Minkowski geometry, or what is the
same by the world function \cite{S60}

\begin{equation}
\sigma _{{\rm M}}\left( x,x^{\prime }\right) =\sigma _{{\rm M}}\left( t,{\bf %
x},t^{\prime },{\bf x}^{\prime }\right) =\frac 12\left( c^2\left(
t-t^{\prime }\right) ^2-\left( {\bf x-x}^{\prime }\right) ^2\right)
\label{a1.1}
\end{equation}
where $c$ is the speed of the light, $x=\left\{ t,{\bf x}\right\} $ and $%
x^{\prime }=\left\{ t^{\prime },{\bf x}^{\prime }\right\} $ are coordinates
of two arbitrary points in the event space.

The second problem is the problem of stochastic microparticles motion, which
cannot be understood and explained in the scope of deterministic classical
physics. To describe phenomena connected with the stochastic microparticle
motion, one should modify the space-time geometry in addition. One should
substitute the world function $\sigma _{{\rm M}}$ by $\sigma $ 
\begin{equation}
\sigma \left( x,x^{\prime }\right) =\sigma _{{\rm M}}\left( x,x^{\prime
}\right) +D\left( \sigma _{{\rm M}}\left( x,x^{\prime }\right) \right) ,
\label{a1.2}
\end{equation}
where $\sigma _{{\rm M}}$ is the Minkowski world function (\ref{a1.1}), and 
\begin{equation}
D=D(\sigma _{{\rm M}})=\left\{ 
\begin{array}{ccc}
d &  & \sigma _{{\rm M}}>\sigma _0 \\ 
0 &  & \sigma _{{\rm M}}\leq 0
\end{array}
\right. ,  \label{a1.3}
\end{equation}
\[
\ d=\frac \hbar {2bc}={\rm const}\approx 10^{-21}{\rm cm}^2, \qquad \sigma
_0={\rm const}\approx d\approx 10^{-21}{\rm cm}^2 
\]
is a correction, called distortion. Here $\hbar $ is the quantum constant, $%
b\approx 10^{-17}{\rm g/cm}$ is a new universal constant. Values of
distortion $D$ within $[0,\sigma _0]$ are yet unknown. They are to be
established as a result of further investigations.

Formally the modification of geometry is very slight, as far as the
distortion $D$ is a small correction to the Minkowski world function.
Nevertheless at the transition from (\ref{a1.1}) to (\ref{a1.2}) the
space-time model changes qualitatively no less, than at the transition from
Newtonian model to the Minkowski one. The space-time geometry, generated by
the world function (\ref{a1.2}), is not a Riemannian geometry. We shall
refer to it as T-geometry. The T-geometry is nondegenerate geometry. It
means that at any point of the space-time there exists many unit timelike
vectors parallel to a given timelike vector, and motion of free particles is
stochastic, although the T-geometry in itself is deterministic. It seems
rather evident that the free particle motion in the space-time with
stochastic geometry is stochastic\cite{M42,W44,M51,SS60,B70,B71}, but a
stochastic motion of a free particle in the deterministic space-time looks
rather unexpected and needs an explanation. It will be given in the second
section.

The classical physics in the event space with T-geometry (\ref{a1.2})
explains phenomena, conditioned by the stochasticity of microparticle motion
(known as quantum effects) freely and without any additional suppositions or
hypotheses. The supposition (\ref{a1.2}) on the character of geometry is not
an additional hypothesis. It is simply a correction of the Minkowski
geometry, which is used instead of (\ref{a1.1}). In other words, the
relation (\ref{a1.2}) is a hypothesis in the same degree, as the statement,
that the space-time geometry is the Minkowski one, is a hypothesis. As one
can see from (\ref{a1.2}), (\ref{a1.3}) the world function $\sigma $ differs
essentially from $\sigma _{{\rm M}}$ only for small space-time intervals of
the order $10^{-10}{\rm cm}$, i.e. in the microcosm.

In the beginning of the XX century the T-geometry was not known for a number
of reasons, which will be discussed in the second section, and the second
problem was solved differently. The Newtonian space-time model was
conserved, but a number of additional hypotheses on the microparticle motion
laws was taken. These additional hypotheses are known as quantum mechanics
principles. The conception of such a solution of the microparticle
stochasticity problem is called the quantum mechanics (QM).

The quantum mechanics is a non-relativistic theory from outset, i.e. the
first problem and the second one are solved separately. Thereafter the
problem of unification of quantum mechanics (QM) with the relativity theory
(RT) arises. The scheme of the conventional research program for microcosm
investigation looks as follows 
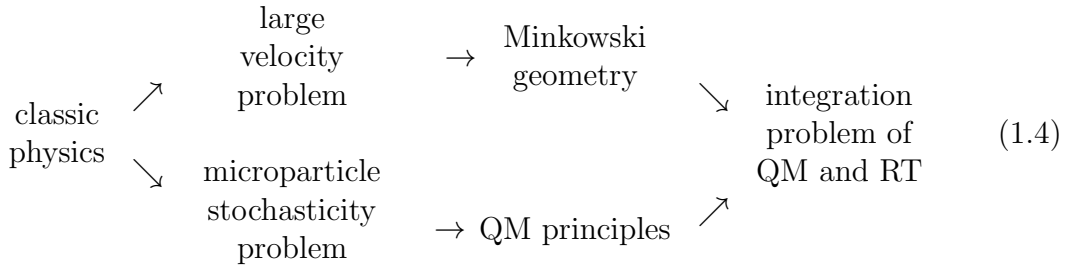
\begin{equation}
\begin{array}{c}
\mbox{classic} \\ 
\mbox{physics}
\end{array}
\begin{array}{c}
\nearrow \\ 
\\ 
\searrow
\end{array}
\begin{array}{c}
\begin{array}{c}
\mbox{large} \\ 
\mbox{velocity} \\ 
\mbox{problem}
\end{array}
\\ 
\\ 
\begin{array}{c}
\mbox{microparticle} \\ 
\mbox{stochasticity} \\ 
\mbox{problem}
\end{array}
\end{array}
\begin{array}{c}
\begin{array}{c}
\rightarrow
\end{array}
\begin{array}{c}
\mbox{Minkowski} \\ 
\mbox{geometry}
\end{array}
\\ 
\\ 
\\ 
\\ 
\begin{array}{c}
\rightarrow
\end{array}
\mbox{QM principles}
\end{array}
\begin{array}{c}
\\ 
\searrow \\ 
\\ 
\\ 
\nearrow
\end{array}
\begin{array}{c}
\mbox{integration} \\ 
\mbox{problem of} \\ 
\mbox{QM and RT}
\end{array}
\label{a1.4}
\end{equation}
The quantum relativistic field theory and elementary particle theory are
different sides of the problem of unification QM with RT.

Unprejudiced observer known another possible solution (\ref{a1.2}) of the
problem is inclined to interpret this in the sense, that the research
program (\ref{a1.4}) uses untrue assertion (inadequate space-time model)
which manifests itself as contradictions appearing in different places of
theory. One needs to introduce new hypotheses, compensating specific
manifestations of the untrue assertion of the theory. At the same time
researchers developing the theory are inclined to connect all arising
problems with complexity of physical phenomena in microcosm. Such a
situation took place in the science history. It is the Ptolemaic doctrine,
using untrue assertion on the place of the Earth in the center of universe.
Ptolemy and his successors succeeded to describe correctly heavenly bodies
motion in spite of the untrue assertion on the place of the Earth in the
center of universe. These results had come about through use of new
additional hypotheses, compensating original untrue assertion on the place
of the Earth. In spite of success in explanation of astronomical
observations the Ptolemaic doctrine lead to blind alley finally, and the
most reasonable way of overcoming all problems was a substitution of untrue
assertion by the true one.

Something like that is observed in the solution of the integration problem
of QM and RT. One uses inadequate space-time model, and this is an origin of
many problems of contemporary quantum theory. A use of the adequate
space-time model (\ref{a1.2}) removes the integration problem of QM and RT,
because the conception (\ref{a1.2}) is relativistic and quantum originally
(it is quantum in the sense that it includes the quantum constant $\hbar $).
Besides it does not contain any additional hypotheses and principles. The
integration problem and problem of concordance of different principles do
not exist at all. Instead there exists the problem of statistical
description of stochastic microparticles. It is a very serious problem,
because the probabilistic statistical description, used in the
nonrelativistic statistical physics, is ineligible for description of
stochastic relativistic motion. (see detail in sec. 3)

As a whole the scheme of the alternative research program for the microcosm
investigation looks as follows. 
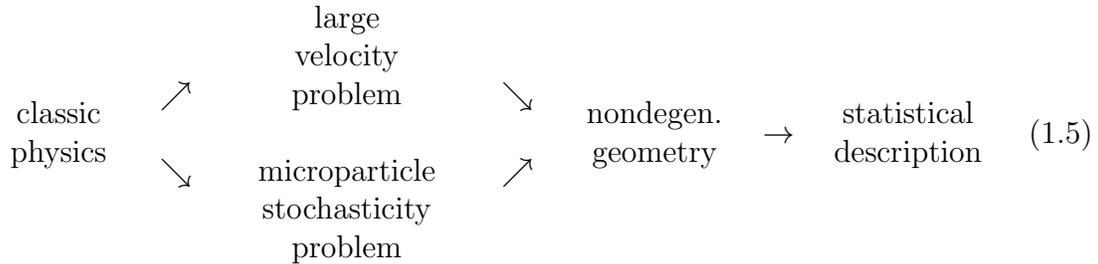
\begin{equation}
\begin{array}{ccccccc}
\begin{array}{c}
\mbox{classic} \\ 
\mbox{physics}
\end{array}
& 
\begin{array}{c}
\nearrow \\ 
\\ 
\searrow
\end{array}
& 
\begin{array}{c}
\begin{array}{c}
\mbox{large} \\ 
\mbox{velocity} \\ 
\mbox{problem}
\end{array}
\\ 
\\ 
\begin{array}{c}
\mbox{microparticle} \\ 
\mbox{stochasticity} \\ 
\mbox{problem}
\end{array}
\end{array}
& 
\begin{array}{c}
\searrow \\ 
\\ 
\nearrow
\end{array}
& 
\begin{array}{c}
\mbox{nondegen.} \\ 
\mbox{geometry}
\end{array}
& \rightarrow & 
\begin{array}{c}
\mbox{statistical} \\ 
\mbox{description}
\end{array}
\end{array}
\label{a1.5}
\end{equation}

The research program (\ref{a1.4}) is a compensating, or Ptolemaic
conception, as far as it uses inadequate space-time model and quantum
principles, compensating inadequacy of this model. One cannot say that this
research program is untrue, because it explains the observed physical
phenomena and yet this program is not a rigorous physical theory, because it
uses inadequate space-time model. Now this program is developed in details,
because several generations of researchers had been working with this
program for the last century. Nevertheless this detailed development does
not prevent from appearance of new problems, which need new hypotheses for
their solution. In this relation the program (\ref{a1.4}) reminds the
Ptolemaic doctrine. We mark this circumstance, referring to the program as
''Ptolemy-2''. Further such a defect of scientific conception will be marked
by a special term ''ptolemyness''. Ptolemyness of the conventional research
program (\ref{a1.4}) is not evident. It becomes clear only after appearance
of alternative research program, which is not Ptolemaic.

Unlike the program (\ref{a1.4}), the research program (\ref{a1.5}) is a
rigorous physical theory. It is not a theory of Ptolemaic type, because it
does not use any additional (compensating) hypotheses besides those which
were used in classical physics. This program is very simple and reasonable.
Now it is very young and slightly developed. All this associates with the
Copernican doctrine at the time of its appearance. We mark this circumstance
referring to the program (\ref{a1.5}) as Copernicus-2. These names have been
given to mark qualitative difference between the two conception Ptolemy-2
and Copernicus-2 and to underline that, analyzing and evaluating interplay
between them, one cannot use the criteria, obtained at the work with
Ptolemaic (compensating) conceptions.

The fact is that that on one hand the Ptolemaic conception, i.e. a theory
containing untrue assertion and compensating it by means of additional
hypotheses, is not a rigorous scientific theory. On the other hand, it is
applied very wide, and researchers of microcosm were forced to work with
different types of Ptolemaic conceptions. As a result one derived the rules
of work and criteria of estimations of obtained results, which are suitable
for work with Ptolemaic conceptions. But these rules and criteria are not
effective with the work with rigorous scientific conceptions, which do not
contains mistakes and inadequate assertions. For work with Ptolemaic
conceptions one needs a ''short logic'', i.e. one uses hypotheses and tries
to make such conclusions from them which could be quickly verified
experimentally. The long chains of logical considerations which cannot be
quickly verified by experiment seem to be doubtfull. Primacy of experiment
over logic and principles is another peculiarity of a Ptolemaic conception,
when any principles are suitable, provided one can explain experimental date
by their use. In many cases a researcher, working only with Ptolemaic
theories, have not enough experience of work with rigorous scientific
theories and cannot evaluate them correctly.

As an example let us consider the following situation. There are two
alternative Ptolemaic conceptions $A$ and $B$. Let the conception $A$ appear
earlier and be accepted by the scientific community. After appearance of the
conception $B$ the proponents of the conception $A$ evaluate the conception $%
B$ as follows. One considers whether the conception $B$ explains the
observed phenomena, which can be explained by the conception $A$. If no, the
conception $B$ is worse. If the conception $B$ explains all phenomena, which
are explained by the conception $A$ and besides it explains some phenomena
which cannot be explained by the conception $A$, then one should prefer the
conception $B$. Finally, if the conception $B$ explains only those
phenomena, which are explained by the conception $A$ and nothing except for
them, one should prefer the conception $A$, because it appeared earlier. The
second conception, leading to the same results, is considered to be
superfluous. Analysis of hypotheses, used in conceptions $A$ and $B$, number
of them and their quality is considered to be superfluous. All this is
valid, provided both conceptions $A$ and $B$ are Ptolemaic. If the
conception $B$ pretends to being rigorous theory, but not Ptolemaic one,
i.e. it contains essentially less additional hypotheses, than the conception 
$B$, one should prefer the conception $B$ even in the case, when it does not
explain nothing besides those phenomena, which are explained by the
conception $A$. In this case one should use the conception $B$, because it
is more promising, and after its development it will explain many phenomena,
which it could not explain at the point in time of appearance.

It is this case that took place in the conflict between the Ptolemy's
doctrine and Copernican one. In first time after appearance the Copernican
doctrine explained nothing in the heavenly bodies motion that cannot be
explained by the Ptolemy's doctrine. Further development of the Copernican
doctrine leads to such results which cannot be imagined by proponents of
Ptolemy. The Copernican doctrine was much simpler, because it did not use
additional (compensating) assertions. It was the simplicity of the
Copernican doctrine, conditioned by its rigor (''non-ptolemyness''), appears
to be the main factor, providing its victory.

The research program Copernicus-2 pretends to the role of rigorous
(non-Ptolemaic) conception, because it does not use the quantum principles
for explanation of non-relativistic quantum effects. On the other hand,
evaluation of the program Ptolemy-2 as a Ptolemaic conception is based on
the fact that there exists the program Copernicus-2, which uses essentially
less number of base assertions and which does not uses, in particular, QM
principles. As for relativistic quantum effects, the program Copernicus-2
cannot say anything about them due to its insufficient development. It
should keep in mind that the program Copernicus-2 is very young, whereas the
program Ptolemy-2 has been developed by several generations of researchers
in the course of several decades. But the program Copernicus-2 promises some
progress in explanation of relativistic quantum effects as a result of
further development, whereas the program Ptolemy-2 does not promise a
progress. Experience of work with Ptolemy-2 shows that at its development
the number of problems increases, and the conception becomes more and more
complex and tangled.

The program Copernicus-2 cannot be considered to be a quite new conception.
All its stages have been known since the beginning of XX century. The fact
that quantum effects can be explained as a result of statistical description
of microparticle stochastic behavior seems very reasonable for many
researchers \cite{M49,F52}. Such an explanation seem to be very plausible in
the light of the success of the statistical physics, which explains the
nature of heat and thermal phenomena in such a way. The fact that the
space-time geometry can be a reason of stochasticity is not new also \cite
{M42,W44,M51,SS60,B70,B71}.

Difficulties of work with the program Copernicus-2 are connected with
insufficient development of geometry and of the conception of the
statistical description. In other words, there were no mathematical tools
which should be sufficiently effective for description of microparticle
stochastic motion. One was forced to construct these mathematical tools,
developing a new conception of geometry and a new conception of statistical
description. It is the development of these new conceptions, that allowed
one to formulate and substantiate the research program Copernicus-2. Thus,
at the development of the research program Coperncus-2 the technical and
mathematical results are main and determining.

The newly developed conceptions of geometry and statistical description are
more general, than existing before. From formal viewpoint the lager
generality is achieved at the cost of reduction of the number of fundamental
concepts, i.e. concepts used at the construction of the conception. In
particular, in T-geometry the concept of a curve is not used, and in the
dynamic conception of statistical description the concept of probability and
that of probability density are not used.

In the second section the new conception of geometry is considered. The most
attention is concentrated on conceptual problems, in particular, one
investigates, how stochastic motion of free particles can appear in the
space-time with deterministic geometry and what is the reason why such a
simple and necessary construction as T-geometry has not been constructed
earlier. In the third section one considers conceptual problems of
statistical description -- a new conception of statistical description
restricted by no constraints of the probability theory. In the fourth
section the dynamic conception of statistical description is applied to the
description of quantum-stochastic particle.

\section{Metric conception of geometry}

Usually a geometry is constructed on the basis of linear space, where linear
operations on vectors are defined. Vector (the main object of the linear
space) is determined by two points: origin of the vector and its end. It is
supposed that in the linear space the origins of all vectors coincide, and
any vector is determined single-valuedly by the point which determines its
end. After definition of the scalar product the linear space turns to vector
Euclidean space. As far as there are one-to-one correspondence between the
vectors of the linear space and points representative their end, the vector
Euclidean space generates the point Euclidean space, where the main
characteristic is the distance $d$ between two points or the world function $%
\sigma =\frac 12 d^2$. The scalar product in the vector Euclidean space is
connected single-valuedly with the world function in the corresponding point
Euclidean space. The scalar product in the vector Euclidean space determines
the world function in the corresponding point Euclidean space, and vice
versa.

Introduction of the linear space as a basis for construction of the
Euclidean space is possible only in the continuous homogeneous space, where
all points and all connections between them are similar. If the continuity
of the space is violated, for instance, removing one point of it, the space
stops to be linear space, because now linear operations are not defined
properly. They lead to a definite result not always. In inhomogeneous space
one has to introduce tangent linear space at any point, and this set of
linear spaces forms a basis for construction of inhomogeneous (Riemannian)
geometry.

The practical work with the event space, considered to be the Minkowski
space, suggests that the geometry is determined by the world function
(distance between any two points) of the event space and that the linear
space is not a necessary attribute of geometry. It plays a role of some
subsidiary construction, which is used for building of geometry and which
can be removed after the geometry has been constructed. If it is really so,
the geometry can be constructed without referring to a linear space. It may
appear that some restrictions, imposed usually on geometry, are generated by
the properties of the linear space, which is used at the construction of
geometry, but not at the geometry itself.

Construction of a geometry, based only on information, contained in the
world function, will be referred to as metric conception of geometry. This
approach is well known as metric geometry \cite{T59,ABN86,BGP92}, But one
did not succeed to carry out it consequently (i.e. without invoking
additional information) and to construct a geometry which should be as
informative as the Euclidean one. One succeeded for the first time to make
this in the papers \cite{R90,R01}.

The idea of the geometry construction on the basis of only world function $%
\sigma$ is very simple. All relations of Euclidean geometry are written in
terms of the world function and declared to be valid for any world function,
i.e. for any geometry. Practically it is important to represent in terms of
world function only the scalar product, because all remaining relations are
expressed finally through it. It is important also not to use the concept of
a curve, defined as a continuous mapping of a segment of real axis on the
space $\Omega$ 
\begin{equation}
{\cal L}:\quad [0,1]\rightarrow \Omega .  \label{b.3}
\end{equation}

Let $\Omega $ be a set of points with the world function $\sigma $, given on 
$\Omega \times \Omega $ 
\begin{equation}
\sigma :\qquad \Omega \times \Omega \rightarrow {\Bbb R}  \label{a2.1}
\end{equation}
\begin{equation}
\sigma \left( P,Q\right) =\sigma \left( Q,P\right) ,\qquad \sigma \left(
P,P\right) =0,\qquad \forall P,Q,\in \Omega  \label{a2.1a}
\end{equation}
Let the totality $V=\{\sigma ,\Omega \}$ be called $\sigma $-space. Vector $%
\overrightarrow{PQ}\equiv {\bf PQ}$ is an ordered set of two points $\left\{
P,Q\right\} $ (point $P$ is an origin of the vector and $Q$ is its end). The
length $|\overrightarrow{PQ}|$ of the vector is determined by the relation $|%
\overrightarrow{PQ}|=\sqrt{2\sigma \left( P,Q\right) }$. The scalar product
of two vectors $\overrightarrow{P_0P_1}$, $\overrightarrow{P_0P_2}$, having
a common origin, is given by the relation 
\begin{equation}
\left( \overrightarrow{P_0P_1}.\overrightarrow{P_0P_2}\right) =\sigma \left(
P_0,P_1\right) +\sigma \left( P_0,P_2\right) -\sigma \left( P_1,P_2\right) ,
\label{a2.2}
\end{equation}
It represents a formula of the cosine theorem for the triangle with vertices
at points $P_0,P_1,P_2$, written in terms of the world function $\sigma $.
The relation (\ref{a2.2}) may be interpreted as a definition of the scalar
product, made without a reference to linear space. To stress independence on
the linear space, the difinition (\ref{a2.2}) will be referred to as the
scalar $\sigma $-product.

Note that the scalar $\sigma$-product can be determined for vectors $%
\overrightarrow{P_0P_1}$, $\overrightarrow{Q_0Q_1}$, having different
origins. In this case the relation (\ref{a2.2}) takes the form 
\begin{equation}
\left( \overrightarrow{P_0P_1}.\overrightarrow{Q_0Q_1}\right) =\sigma \left(
P_0,Q_1\right) +\sigma \left( Q_0,P_1\right) -\sigma \left( P_0,Q_0\right)
-\sigma \left( P_1,Q_1\right) ,  \label{a2.2a}
\end{equation}

Dimension of the space is its another important property, determined by the
maximal number of linearly independent vectors. For $n$ vectors $%
\overrightarrow{P_0P_i}\;\;\;i=1,2,...n$ of the Euclidean space were
linearly independent, it is necessary and sufficient that the Gram's
determinant vanishes 
\begin{equation}
F_n\left( {\cal P}^n\right) =0,\qquad {\cal P}^n\equiv \left\{
P_0,P_1,\ldots P_n\right\}\subset\Omega ,  \label{a2.3}
\end{equation}
where 
\begin{equation}
F_n\left( {\cal P}^n\right) \equiv\det \left\| \left( \overrightarrow{P_0P_i}%
.\overrightarrow{P_0P_k}\right) \right\| ,\qquad i,k=1,2,\ldots n
\label{a2.4}
\end{equation}
It follows from (\ref{a2.2}) and (\ref{a2.4}), that linear independence of
vectors can be defined in terms of the world function without a reference to
linear space.

There exist necessary and sufficient conditions that the $\sigma$-space $%
V=\{\sigma ,\Omega \}$ is $n$-dimensional Euclidean space. They state that
there exists a set of $(n+3)$-point $\sigma$-subspaces $\{\sigma ,{\cal P}%
^{n+2}\}\subset V$, whose world function satisfies some relations. $\sigma$%
-subspaces of this set have $n+1$ common fixed points ${\cal P}^n$. Other
two points $P_{n+1}$, $P_{n+2}$ are arbitrary points of $V$ and running
points of $\sigma$-subspaces $\{\sigma ,{\cal P}^{n+2}\}$ of this set.
Corresponding theorem was proved in \cite{R01}.

It follows from the theorem that information, contained in the world
function, is sufficient for construction of rigorous geometry which is as
rich in content as the Euclidean geometry. Any choice of the world function,
satisfying the condition (\ref{a2.1}), corresponds to some geometry. This
choice is restricted neither continuity condition, nor condition of geometry
degeneracy.

All known geometries (Riemannian, Euclidean) are degenerate geometries.
Non-degenerate geometry is a new type of geometry, and the concept of
degeneracy merits a special discussion. Two vectors $\overrightarrow{P_0P_1}$
and $\overrightarrow{P_0R}$, having common origin are called collinear $%
\overrightarrow{P_0P_1}||\overrightarrow{P_0R}$, if they are linearly
dependent, i.e. if they satisfy the condition 
\begin{equation}
F_2(P_0,P_1,R)=\left| 
\begin{array}{cc}
\left( \overrightarrow{P_0P_1}.\overrightarrow{P_0P_1}\right) & \left( 
\overrightarrow{P_0P_1}.\overrightarrow{P_0R}\right) \\ 
\left( \overrightarrow{P_0R}.\overrightarrow{P_0P_1}\right) & \left( 
\overrightarrow{P_0R}.\overrightarrow{P_0R}\right)
\end{array}
\right| =0,  \label{a2.5}
\end{equation}
which can be written in the form 
\begin{equation}
\cos ^2\vartheta =\frac{\left( \overrightarrow{P_0P_1}.\overrightarrow{P_0R}%
\right)^2 }{\left| \overrightarrow{P_0P_1}\right| ^2\left| \overrightarrow{%
P_0R}\right| ^2}=1  \label{a2.4a}
\end{equation}
The last relation means that the angle $\vartheta $ between vectors is equal
to either $0$, or $\pi $.

Let vector $\overrightarrow{P_0P_1}$ be given in $n$-dimensional Euclidean
space, and $\overrightarrow{P_0R}$ is a vector collinear to $\overrightarrow{%
P_0P_1}$. Then the set ${\cal T}_{P_0P_1}$ of points $R$ 
\begin{equation}
{\cal T}_{P_0P_1}=\left\{ R|F_2\left( P_0,P_1,R\right) =0\right\}
\label{a2.6}
\end{equation}
is a straight line, passing through the points $P_0,P_1$, or, what is the
same, it is a straight line, passing through the point $P_0$, parallel to
vector $\overrightarrow{P_0P_1}$. On the other hand, at the arbitrary world
function the set ${\cal T}_{P_0P_1}$, determined by one equation, describes,
generally, $(n-1)$-dimensional surface. The fact, that in the case of
Euclidean space this $(n-1)$-dimensional surface degenerates to
one-dimensional line, is connected with the special form of the world
function of the Euclidean space. Even small change of the world function
either removes degeneration, and the one-dimensional line turns to hallow $%
(n-1)$-dimensional tube, enveloping the straight, or increases degeneration,
and the one-dimensional line degenerates to two points $P_0,P_1$. Thus, in
the non-degenerate geometry the straights are substituted by hallow tubes.
This fact justifies the name of geometry -- tubular geometry, or briefly
T-geometry.

If there is no continuous coordinate system on the set $\Omega$, it is
difficult to determine whether the set (\ref{a2.6}) is a one-dimensional
line. In this case for estimation of the degeneracy degree one can consider
intersection between the tube ${\cal T}_{P_0P_1}$ and the sphere of radius $%
r=\sqrt{2\sigma(P_0,Q)}$, which passes through the point $Q$ and has its
center at the point $P_0$ 
\begin{equation}
{\cal S}\left( P_0,Q\right) =\left\{ R|\sigma \left( P_0,Q\right) =\sigma
\left( P_0,R\right) \right\}  \label{a2.7}
\end{equation}
In the case of Euclidean space the intersection ${\cal T}_{P_0P_1}\cap {\cal %
S}\left( P_0,Q\right)$ consists of two points $Q_1,Q_2$. The vector $%
\overrightarrow{P_0Q_1}$ is parallel to the vector $\overrightarrow{P_0P_1}$%
, $(\overrightarrow{P_0Q_1}\uparrow \uparrow \overrightarrow{P_0P_1})$, and
vector $\overrightarrow{P_0Q_2}$ is antiparallel to the vector $%
\overrightarrow{P_0P_1}$, $(\overrightarrow{P_0Q_2}\downarrow \uparrow 
\overrightarrow{P_0P_1})$.

In other words, at the degenerate geometry at any point $P_0$ there is only
one vector of given length, which is parallel to the given vector $%
\overrightarrow{P_0P_1}$, and only one vector of given length, which is
antiparallel $\overrightarrow{P_0P_1}$.

In the case of non-degenerate geometry the intersection ${\cal T}%
_{P_0P_1}\cap {\cal S}\left( P_0,Q\right)=\omega _{+}\cup \omega _{-}$ is
divided into two such subsets $\omega _{+}$ 
\"{}%
$\omega _{-}$, that the points $Q_1\in \omega _{+}$ determine vectors $%
\overrightarrow{P_0Q_1}$, $\overrightarrow{P_0Q_1}\uparrow\uparrow 
\overrightarrow{P_0P_1} $, and points $Q_2\in \omega _{-}$ determine vectors 
$\overrightarrow{P_0Q_2}$, $\overrightarrow{P_0Q_2}\downarrow \uparrow 
\overrightarrow{P_0P_1}$. Each of subsets $\omega _{+}$ and $\omega _{-}$
contains many points. This corresponds to the fact that in the
non-degenerate geometry at any point there are many vectors of given length $%
r=\sqrt{2\sigma(P_0,Q)}$, which are parallel (antiparallel) to the given
vector $\overrightarrow{P_0P_1}$.

Non-degeneracy of the space-time geometry, i.e. existence of many timelike
vectors of fixed length parallel to a given timelike vector at any point, is
a reason of the free particle stochastic motion. To show this, let us
consider the event space, where at any point $P_0$ there are many timelike
vectors ${\bf P}_0{\bf P}_1$ of the given length $|{\bf P}_0{\bf P}_1|=\mu $%
, parallel to the given timelike vector ${\bf P}_0{\bf Q}_1$. Note that in
the Minkowski geometry there is only one timelike vector ${\bf P}_0{\bf P}_1$
of the given length, parallel to timelike vector ${\bf P}_0{\bf Q}_1$.

In the Minkowski space-time the particle world line can be approximated by a
broken line, consisted of rectilinear links of the same length. Then the
joining points $\ldots P_{i-1},P_i,P_{i+1}$ are such, that the vector ${\bf P%
}_i{\bf P}_{i+1}$ is proportional to the particle momentum, and its length $|%
{\bf P}_i{\bf P}_{i+1}|=\mu $ is proportional to its mass $m=b\mu $, where $%
b\approx 10^{-17}\mbox{g/cm}$ is some universal constant and $\mu $ is
geometric particle mass. If the particle is free, according to the Galilean
law of inertia the adjacent links are parallel, i.e. the vector ${\bf P}_i%
{\bf P}_{i+1}$ is parallel to the vector ${\bf P}_{i+1}{\bf P}_{i+2}$, $%
i=0,\pm 1,\pm 2,\ldots $.

Let us define the world line of a free particle as a broken line with
parallel links. Then in the Minkowski space-time position of all links is
determined single-valuedly, provided one fixes position of one link.
Determinism of the broken line means determinism of the particle world line,
what conditions determinism of the free particle motion. In T-geometry,
where there are many vectors, parallel to the given one, fixing of a
position of one link does not lead to single-valued determination of the
remaining links position. It means that in such a space-time the free
particle motion is stochastic, although the geometry in itself is
deterministic.

In general, the T-geometry is non-Riemannian geometry. In some cases, when
the set of vectors of fixed length, parallel to a given vector, degenerates
into one vector, T-geometry degenerates into a Riemannian geometry. For
instance, being a pseudo-Riemannian geometry, the Minkowski geometry is a
special case of T-geometry.

Thus, T-geometry is rather general construction, having such an important
property as non-degeneracy. The non-degeneracy of a geometry is a new
unknown earlier property of geometry. Importance of this property is
comparable with such important properties of geometry as continuity and
homogeneity. It seems rather enigmatic, why such a simple and general
construction as T-geometry was not known earlier. Why was such a property of
geometry as non-degeneracy not known before the end of XX century? Absence
of T-geometry in the list of possible geometries does not allow to solve
correctly the microparticle stochasticity problem.

Absence of T-geometry at the beginning of the XX century even in the form of
a speculative construction is explained, apparently, by existence of a
discriminator, used at the geometry construction. The point is that,
constructing geometry in terms of some fundamental concepts (for instance,
such as dimension, coordinate system, distance, curve, etc.), one
discriminates automatically those geometries, which are incompatible with at
least one of these fundamental concepts. For instance, the Cartesian
coordinate system is a discriminator of inhomogeneous (Riemannian) geometry.
That is the reason why a Riemannian geometry is constructed in arbitrary
(not Cartesian) coordinate system with all its attributes in the form of
Christoffel symbols and covariant derivatives. If one declares that a
Riemannian geometry is described in the Cartesian coordinate system, where
the metric tensor $g_{ik}=$const, the nonhomogeneity of geometry is
discriminated, and only homogeneous (Euclidean) geometry remains. In XIX
century the Cartesian coordinate system was considered as something immanent
to geometry in itself, and apparently, this circumstance stipulates
prejudice of many mathematicians of XIX century against the Riemannian
geometry.

The concept of a curve is a discriminator of non-degenerate geometries. This
fact was realized quite recently \cite{R001}. One attempted to generalize
the Riemannian and metric geometries. One attempted to generalize the metric
geometry, removing the triangle axiom. Such a geometry is referred to as
distant geometry. K. Menger \cite{M28} and L. Blumenthal \cite{B53}
attempted to construct distant geometry. But metric geometry, or distant
geometry, constructed without a use of the concept of a curve appears to be
very poor geometries, because they contained few geometrical objects. To
obtain more rich in content geometry, one uses the concept of a curve.
Essentially this discriminates any possibility of an effective application
of the triangle axiom remove, and a non-degenerate geometry cannot appear.

Thus, on the one hand, at construction of a geometry a use of the concept of
the curve discriminates its non-degeneracy automatically. On the other hand,
the concept of the curve is necessary for constructing geometrical objects,
and it is not clear, what can substitute this very important concept of
Riemannian geometry. Now the most of mathematician consider the concept of
the curve (\ref{b.3}) as a necessary attribute of any geometry. This is an
origin of their prejudice against the T-geometry, and reminds the situation
of the end of XIX century, when, considering the Cartesian coordinate system
to be an attribute of any geometry, the most of mathematician had prejudice
against the Riemannian geometry.

In the Riemannian geometry the concept of a continuous curve has two base
functions: (1) the curve is a fundamental concept, used at construction of a
geometry, (2) the curve is a tool for construction of geometrical objects.
Geometrical object is some set ${\cal O}\subset \Omega $ of points. Usually
it is a continual set. In T-geometry all geometrical relations are expressed
via the world function and the first function of the curve appears to be not
claimed.

The second function of the curve is used in the Riemannian geometry, where a
geometrical object is build usually as a trace of motion of a more simple
geometrical object. For instance a one-dimensional curve ${\cal L}$ is
considered to be a trace of a moving point. It is described by the
continuous mapping (\ref{b.3}). A two-dimensional surface ${\cal S}$ is
considered to be trace of moving one-dimensional curve. It is described by a
continuous mapping 
\begin{equation}
{\cal S}:\qquad [0,1]\times [0,1]\rightarrow \Omega  \label{b1.2}
\end{equation}
etc. Such a construction of a geometrical object contains a continuous
mapping of the type continuum $\to $ continuum, which is very difficult for
investigations, because before investigations of such mappings one needs at
least to label them. But even the problem of labelling of all possible
mappings of the type continuum $\to $ continuum is very complicated because
of large power of the set of such mappings.

To investigate mappings of such a kind and to use them in geometry, one
needs to separate only small part of them, imposing constraints on
properties of the space $\Omega $ (for instance such constraints as
continuity and topological properties). These constraints reduce the
geometry generality in incontrollable way.

In T-geometry a geometrical object ${\cal O}$ is described by means of the
skeleton-envelope method. Any geometrical object ${\cal O}$ is considered to
be a set of intersection and joins of elementary geometrical objects (EGO).

Elementary geometrical object ${\cal E}$ is described as a set of zeros of
some function 
\begin{equation}
f_{{\cal P}^n}:\qquad \Omega \rightarrow {\Bbb R},\qquad {\cal P}^n\equiv
\left\{ P_0,P_1,...P_n\right\} \subset \Omega  \label{b1.4}
\end{equation}
It is represented in the form 
\begin{equation}
{\cal E}={\cal E}_f\left( {\cal P}^n\right) =\left\{ R|f_{{\cal P}^n}\left(
R\right) =0\right\}  \label{b1.5}
\end{equation}
The finite set ${\cal P}^n\subset \Omega $ will be referred to as the
skeleton of elementary geometrical object ${\cal E}\subset \Omega $. The
continual set ${\cal E}\subset \Omega $ is referred to as the envelope of
the skeleton ${\cal P}^n$. The function $f_{{\cal P}^n}$, determining the
elementary geometrical object (EGO) is a function of parameters ${\cal P}%
^n\subset \Omega $ and of the running point $R\in \Omega $. The function $f_{%
{\cal P}^n}$ is supposed to be algebraic function of several arguments $%
w=\left\{ w_1,w_2,...w_s\right\} $, $s=(n+2)(n+1)/2$. Each of arguments $w_k$
is the world function $w_k=\sigma \left( Q_k,L_k\right) $ of two arguments $%
Q_k,L_k\in \left\{ R,{\cal P}^n\right\} $, belonging either to the skeleton $%
{\cal P}^n$, or to the running point $R$.

For instance, 
\begin{equation}
{\cal S}(P_0,P_1)=\left\{ R|f_{P_0P_1}\left( R\right) =0\right\} ,\qquad
f_{P_0P_1}\left( R\right) =\sqrt{2\sigma \left( P_0,P_1\right) }-\sqrt{%
2\sigma \left( P_0,R\right) }  \label{b1.6}
\end{equation}
is a sphere, passing through the point $P_1$ and having its center at the
point $P_0$. Ellipsoid ${\cal EL}$, passing through the point $P_2$ and
having the focuses at points $P_0,P_1$ $\left( P_0\neq P_1\right) $ is
described by the relation 
\begin{equation}
{\cal EL}(P_0,P_1,P_2)=\left\{ R|f_{P_0P_1P_2}\left( R\right) =0\right\} ,
\label{b1.7}
\end{equation}
where 
\begin{equation}
f_{P_0P_1P_2}\left( R\right) =\sqrt{2\sigma \left( P_0,P_2\right) }+\sqrt{%
2\sigma \left( P_1,P_2\right) }-\sqrt{2\sigma \left( P_0,R\right) }-\sqrt{%
2\sigma \left( P_1,R\right) }  \label{b1.8}
\end{equation}
If focuses $P_0,P_1$ coincide $\left( P_0=P_1\right) $, the ellipsoid ${\cal %
EL}(P_0,P_1,P_2)$ degenerates into a sphere ${\cal S}(P_0,P_2)$. If the
points $P_1,P_2$ coincide $\left( P_1=P_2\right) $, the ellipsoid ${\cal EL}%
(P_0,P_1,P_2)$ degenerates into a segment of a straight line ${\cal T}%
_{[P_0P_1]}$ between the points $P_0,P_1$. 
\begin{equation}
{\cal T}_{[P_0P_1]}={\cal EL}(P_0,P_1,P_1)=\left\{ R|f_{P_0P_1P_1}\left(
R\right) =0\right\} ,  \label{b1.9}
\end{equation}
\begin{equation}
f_{P_0P_1P_1}\left( R\right) =S_2\left( P_0,R,P_1\right) \equiv \sqrt{%
2\sigma \left( P_0,P_1\right) }-\sqrt{2\sigma \left( P_0,R\right) }-\sqrt{%
2\sigma \left( P_1,R\right) }  \label{b1.10}
\end{equation}
Another functions $f$ generate another envelopes of elementary geometrical
objects for the given skeleton ${\cal P}^n$.

For instance, the set of two points $\{P_0,P_1\}$ forms a skeleton not only
for the tube ${\cal T}_{P_0P_1}$, but also for the segment ${\cal T}%
_{[P_0P_1]}$ of the tube (straight) (\ref{b1.9}), and for the tube ray $%
{\cal T}_{[P_0P_1}$, which is defined by the relation 
\begin{equation}
{\cal T}_{[P_0P_1}=\left\{ R|S_2\left( P_0,P_1,R\right) =0\right\}
\label{a2.9}
\end{equation}
where the function $S_2$ is defined by the relation (\ref{b1.10}).

Any mapping (\ref{b1.4}) of the type continuum$\to $continuum is given and
fixed, because the function $f_{{\cal P}^n}$ is a known function of its
argument and parameters ${\cal P}^n$. Any such function $f_{{\cal P}^n}$
determines some class of elementary geometrical objects (EGO). A set of such
functions is $n$-parametric set of functions. To build and investigate this
class of EGOs, one does not need to impose any constraints on the set $%
\Omega $, or on the world function. Thus, the skeleton-envelope method of
building of geometrical objects deals only with investigation of
comparatively simple mappings of the form 
\begin{equation}
m_n:\qquad I_n\rightarrow {\Bbb R},\qquad I_n=\left\{ 0,1,...,n\right\}
\label{b1.11}
\end{equation}
and it does not need imposition of constraints on the set $\Omega $. Such
mappings are connected with construction and investigation of EGO skeletons.
Investigating a skeleton, one investigates simultaneously corresponding
classes of EGOs, because at the fixed function (\ref{b1.4}) any EGO is
connected rigidly with its skeleton.

Sometimes, investigating a geometrical object, it is sufficient to
investigate its skeleton, which a countable set of points and can be
investigated easier, than the continual set of points, forming the
geometrical object in itself. For instance, analyzing reasons of the free
particle stochasticity, we have restricted ourselves to investigation of the
skeleton $\ldots P_{i-1},P_{i},P_{i+1}, \ldots$ of the broken tube. It
simplifies our analysis essentially.

The skeleton-envelope method simplifies essentially the problem of
geometrical object building. It allows to separate the problem into informal
problem of the skeleton construction and a formal procedure of the envelope
construction, using its skeleton. Taking in to account that the problem of
the envelope construction in accord with its skeleton is formalized, one can
consider the envelope of the geometrical object to be an attribute of its
skeleton.

\section{Dynamic conception of statistical description}

There are numerous attempts of considering the quantum description of
microparticle motion as a result of statistical description of their
stochastic motion \cite{M49,F52}. As a rule they are founded on the
probability theory which is not suitable for description of relativistic
stochastic motion. But stochastic motion, generated by the quantum
stochasticity is relativistic. Inapplicability of the probability theory for
description of relativistic stochastic processes is connected with the fact,
that the concept of probability density supposes a possibility of the event
space separation to sets of simultaneous independent events. It is
impossible in the relativistic theory, where the absolute simultaneity is
absent. Formally this is displayed in the fact, that at the description of
stochastic relativistic particle the object of statistical description is
such a lengthy physical object as world line in the space-time, whereas in
the non-relativistic case the object of the statistical description is the
pointlike particle in the three-dimensional space.

Numerous unsuccessful attempts of representing the quantum mechanics as a
result of the probabilistic statistical description had discredited the idea
in itself to reduce the quantum mechanical description to the statistical
description of randomly moving particles. Now many serious researchers
consider sceptically a possibility of the quantum mechanics reduction to the
statistical description of stochastically moving particles, although the
quantum mechanics is considered to be a statistical theory.

Strictly, the term ''statistical description'' means a description,
containing many similar objects, a reference to a probability concept or
probability density being unnecessary. Moreover, such a reference is
undesirable, as far as the statistical description, founded on the concept
of probability, is restricted by a possibility of the probability
introduction. Dynamic conception of statistical description seems to be more
effective, although it is less informative. Essence of the dynamic
conception of statistical description is formulated as follows \cite
{R80,R991}.

Let ${\cal S}_{{\rm st}}$ be a stochastic system, i.e. dynamic system%
\footnote{%
Conventional terminology contains only terms "stochastic system" and
"dynamic system". The concept collective with respect to the two terms is
absent. For this reason the term "dynamic system" is used as a collective
term with respect to terms "non-deterministic dynamic system" (instead of
customary "stochastic system") and "deterministic dynamic system" (instead
of customary "dynamic system").}, experiments with which are irreproducible,
and for which dynamic equations do not exist. For instance, let ${\cal S}_{%
{\rm st}}$ be an electron flying through a narrow slit in a diaphragm and
hitting the screen at some point $x_1$. Another experiment, produced with an
electron, prepared in the same way, leads to its hit at another point $x_2$,
which does not coincide with $x_1$, generally. In other words the electron $%
{\cal S}_{{\rm st}}$ is a stochastic system, and experiments with it are
irreproducible.

If one produces $N$, $(N\to \infty )$ experiments with ${\cal S}_{{\rm st}}$%
, the obtained distribution of electrons over the screen can be
reproducible. It can be reproduced in other series of $N_1$, $(N_1\to \infty
)$ experiments. It means that the dynamic system ${\cal E}\left[ N,{\cal S}_{%
{\rm st}}\right] $, consisting of many independent non-deterministic
(stochastic) dynamic systems ${\cal S}_{{\rm st}}$, is a deterministic
system, experiments with which are reproducible, and for which there are
dynamic equations, although dynamic equations do not exist for ${\cal S}_{%
{\rm st}}$. The dynamic system ${\cal E}\left[ {\cal S}\right] ={\cal E}%
\left[ \infty ,{\cal S}\right] $ is known as a statistical ensemble, and
dynamic systems ${\cal S}$, constituting it are referred to as the
statistical ensemble elements. Elements of the ensemble can be deterministic
dynamic systems ${\cal S}_{{\rm d}}$, as well as stochastic ones ${\cal S}_{%
{\rm st}}$. Being a dynamic system, the statistical ensemble ${\cal E}$ may
be an element of other statistical ensemble ${\cal E}^{\prime }$, which in
turn may be an element of the statistical ensemble ${\cal E}^{\prime \prime
} $, etc.

Idea of the dynamic conception of the statistical description lies in the
fact that it is impossible to investigate the stochastic system ${\cal S}_{%
{\rm st}}$, because of irreproducibility of experiments with it, but one can
investigate the statistical ensemble ${\cal E}\left[ N,{\cal S}_{{\rm st}}%
\right] $ as a deterministic dynamic system, and on the basis of these
results one can make some conclusions on the properties of stochastic system 
${\cal S}_{{\rm st}}$.

Why does the set ${\cal E}\left[ {\cal S}_{{\rm st}}\right] $ of many
independent stochastic systems ${\cal S}_{{\rm st}}$ turn to a deterministic
dynamic system? Apparently, because that typical features are summed or
averaged, but random ones compensate themselves. Is this so or not, but it
is evident that ${\cal E}\left[{\cal S}_{{\rm st}}\right] $ is a
deterministic dynamic system, and it is a basis of the statistical
description. In any case one can consider this statement as a principle,
which will be referred to as statistical principle \cite{R80,R991}.

The statistical ensemble have several important properties. Using them, one
can transform statistical description in such a way, that it loses its
statistical features and will be perceived as purely dynamical. Such a
description stops to resemble a statistical description, understood as a
probabilistic statistical description. There are three basic properties of
the statistical description.

\noindent (1) Properties of the statistical ensemble do not depend on the
number $N$ of its elements, if this number is enough large, i.e. $N\to\infty$%
.

\noindent (2) A statistical ensemble may be an element of other statistical
ensemble.

\noindent (3) In the simplest case of pure ensemble ${\cal E}\left[ {\cal S}%
_{{\rm st}}\right] $ of stochastic systems ${\cal S}_{{\rm st}}$ coincides
with the dynamic system ${\cal E}_{{\rm red}}\left[ {\cal S}_{{\rm d}}\right]
={\cal S}\left[ {\cal S}_{{\rm d}}\right] $, consisting of many interacting
deterministic systems ${\cal S}_{{\rm d}}$. The form of interaction of
deterministic systems ${\cal S}_{{\rm d}}$ is determined by the
stochasticity character of stochastic systems ${\cal S}_{{\rm st}}$. This
allows to label the stochasticity character by the form of interaction and
to reduce description of stochasticity to interaction of deterministic
systems.

Let us start from the first property, which admits to normalize the ensemble
state. Let the stochastic system ${\cal S}_{{\rm st}}$ represent a
microparticle, whose state is described by its position ${\bf x}$ and
momentum ${\bf p}$. Then at large enough $N$ the ensemble ${\cal E}\left[ N,%
{\cal S}_{{\rm st}}\right] $ represents a distributed fluidlike dynamic
system. There are an action for such a system ${\cal A}\left[ N,\varphi ,%
{\bf \xi }\right] $, where $\varphi {\bf =}\varphi \left( t,{\bf x}\right) $%
, and ${\bf \xi =\xi }\left( t,{\bf x}\right) $ are dynamic variables,
describing the fluid state. The action for the statistical ensemble has the
property 
\begin{equation}
{\cal A}\left[ aN,\varphi ,{\bf \xi }\right] =a{\cal A}\left[ N,\varphi ,%
{\bf \xi }\right] ,\qquad a=\mbox{const},\qquad a>0.  \label{a3.1}
\end{equation}
It generates dynamic equations and the energy-momentum tensor $T_k^i$.
Besides, for the dynamic system ${\cal E}\left[ N,{\cal S}_{{\rm st}}\right] 
$ one can introduce the particle density $j^0$ and the particle flux density 
$j^\alpha $, $\alpha =1,2,3$. Due to relation (\ref{a3.1}) the ensemble
properties do not depend on the number $N$ of its elements. But one may
consider that this property is fulfilled for any $N$ and, setting formally $%
N=1$, consider an ensemble, consisting of one element. More exactly it means
that, if ${\cal A}\left[ N,\varphi ,{\bf \xi }\right] $ is the action for
the statistical ensemble ${\cal E}\left[ N,{\cal S}_{{\rm st}}\right]$, the
action 
\[
{\cal A}_{{\rm av}}\left[ \varphi ,{\bf \xi }\right] =\lim_{N\rightarrow
\infty } \frac{{\cal A}\left[ N,\varphi ,{\bf \xi }\right] }{N} 
\]
is the action for $\left\langle {\cal S}_{{\rm st}}\right\rangle $. Such a
statistical ensemble will be referred to as average dynamic system $%
\left\langle {\cal S}_{{\rm st}}\right\rangle $. Thus, $\left\langle {\cal S}%
_{{\rm st}}\right\rangle ={\cal E}\left[ N,{\cal S}_{{\rm st}}\right] _{N=1}$%
. The average dynamic system $\left\langle {\cal S}_{{\rm st}}\right\rangle $%
, constructed on the basis of the stochastic system ${\cal S}_{{\rm st}}$,
is a deterministic dynamic system, for which a value of any physical
quantity $q $ can be interpreted as the mean value $\left\langle
q\right\rangle $ of the same quantity $q$ for the stochastic system ${\cal S}%
_{{\rm st}}$. The average dynamic system $\left\langle {\cal S}_{{\rm st}%
}\right\rangle $ is a deterministic system, having dynamic equations. Using
these equations, one can calculate evolution of the mean value $\left\langle
q\right\rangle $ of any physical quantity $q$ for the stochastic system $%
{\cal S}_{{\rm st}}$.

As a result of such approach the statistical description of stochastic
system ${\cal S}_{{\rm st}}$ reduces to consideration of a deterministic
system $\left\langle {\cal S}_{{\rm st}}\right\rangle $, but the
circumstance that $\left\langle {\cal S}_{{\rm st}}\right\rangle $ remains
to be a statistical ensemble may drop out of consideration.

Thus, one can consider simultaneously two dynamic systems ${\cal S}_{{\rm st}%
}$ and $\left\langle {\cal S}_{{\rm st}}\right\rangle $. The system ${\cal S}%
_{{\rm st}}$ is concentrated, but stochastic. The system $\left\langle {\cal %
S}_{{\rm st}}\right\rangle $ is distributed, but deterministic. They cannot
be confused, and one should use different terms and designations for them.
The state of the distributed system $\left\langle {\cal S}_{{\rm st}%
}\right\rangle $ can be described by the wave function $\psi $ (it will be
shown below). It is this system, that is considered usually in quantum
mechanics. It is considered as a dynamic system, describing a real physical
particle. As for the stochastic system ${\cal S}_{{\rm st}}$, it does not
appear in the quantum mechanics technique. It may be disregarded, until one
deals only with dynamics, where only the average dynamic system $%
\left\langle {\cal S}_{{\rm st}}\right\rangle $ appears. But discussing the
measurement processes, such a disregard of the stochastic system ${\cal S}_{%
{\rm st}}$ is inadmissible, because there are several different measurement
procedures, where the systems ${\cal S}_{{\rm st}}$ and $\left\langle {\cal S%
}_{{\rm st}}\right\rangle $ play different roles.

Unfortunately, in quantum mechanics almost never one differs systems ${\cal S%
}_{{\rm st}}$ and $\left\langle {\cal S}_{{\rm st}}\right\rangle $.
Furthermore, considering the measurement process, one uses the same term for
them, what is inadmissible even from viewpoint of usual logic. Besides, at
such an ''generalized terminology'' different measurement procedures merge
into one procedure, which is interpreted by different researchers in
different ways, depending on, which of two systems ${\cal S}_{{\rm st}}$ or $%
\left\langle {\cal S}_{{\rm st}}\right\rangle $ is taken into account at
this time. Numerous paradoxes (wave function collapse, Schr\"odinger cat
paradox, Einstein -- Podolski -- Rosen paradox \cite{EPR35}, etc.) are
corollaries of such a consideration, although in reality there are no
paradoxes. There is only confusion, when the same term is used for two
different objects. Note, that paradoxes arise only at the discussion of the
measurement process, where both systems ${\cal S}_{{\rm st}}$ and $%
\left\langle {\cal S}_{{\rm st}}\right\rangle $ appear. At the discussion of
dynamics, where only the system $\left\langle {\cal S}_{{\rm st}%
}\right\rangle $ appears, there are no paradoxes.

The second property of the statistical ensemble means that one statistical
ensemble may be an element of the other one. Such an organization of a
statistical description is useful in the following relation. If elements of
a statistical ensemble are deterministic dynamic systems ${\cal S}_{{\rm d}}$%
, i.e. such dynamical systems, for which there are dynamic equations, a
construction of dynamic equations for the statistical ensemble ${\cal E}%
\left[ \infty,{\cal S}_{{\rm d}}\right] $ is a formal procedure, which can
be carried out easily, provided dynamic equations for ${\cal S}_{{\rm d}}$
are known. If elements of the statistical ensemble are nondeterministic
dynamic systems ${\cal S}_{{\rm st}}$, i.e. such dynamic systems, for which
there are no dynamic equations, construction of dynamic equations for the
statistical ensemble ${\cal E}\left[ \infty,{\cal S}_{{\rm st}}\right] $ is
a complicated informal procedure.

Let us explain this in an example of a description of deterministic particle 
${\cal S}_{{\rm d}}$, whose motion is described by the Hamilton function $%
H(t,{\bf x},{\bf p},)$, where ${\bf x}=\left\{ x^\alpha \right\} \;\;\alpha
=1,2,...n,$ are generalized coordinates and ${\bf p}=\left\{ p_\alpha
\right\} \;\;\alpha =1,2,...n$ is a generalized momentum. The most general
statistical ensemble ${\cal E}_{{\rm gen}}\left[ {\cal S}_{{\rm d}}\right] $
is described usually by the distribution function $F(t,{\bf x},{\bf p})$,
satisfying the Liouville equation. ${\cal E}_{{\rm gen}}\left[ {\cal S}_{%
{\rm d}}\right] $ may be considered to be a statistical ensemble ${\cal E}_{%
{\rm gen}}\left[ {\cal E}_{{\rm p}}\right] $, whose elements are statistical
ensembles ${\cal E}_{{\rm p}}\left[ {\cal S}_{{\rm d}}\right] $ of special
type, whose elements are dynamic systems ${\cal S}_{{\rm d}}$.

Following von Neumann \cite{N32}, we shall refer to the statistical ensemble
of special type ${\cal E}_{{\rm p}}\left[ {\cal S}\right] $ as a pure
ensemble, because it admits a description in terms of the wave function. (It
will be shown below). By definition the pure statistical ensemble is such a
statistical ensemble ${\cal E}_{{\rm p}}\left[ {\cal S}_{{\rm d}}\right] $,
which is described by the distribution function 
\begin{equation}
F_{{\rm p}}(t,{\bf x},{\bf p})=\rho (t,{\bf x})\delta ({\bf p}-{\bf P}(t,%
{\bf x}))  \label{c3.0}
\end{equation}
It satisfies a system of dynamic equations written for independent variables 
$\{t,{\bf x}\}$, i.e. in the configuration space of coordinates ${\bf x}$.
In other words, the pure statistical ensemble is described in terms of
several functions $\rho (t,{\bf x})$ and ${\bf P}(t,{\bf x})$ of only
argument ${\bf x}$ instead of a description in terms of one function of
arguments ${\bf x},{\bf p}$. The system of dynamic equations for these
functions is derived as a result of the substitution (\ref{c3.0}) into the
Liouville equation for the distribution function $F(t,{\bf x},{\bf p})$ and
subsequent integration with respect to variable ${\bf p}$ with the weight
multipliers $1$ and ${\bf p}$.

Now if the particle is a stochastic one ${\cal S}_{{\rm st}}$, an informal
procedure is only construction of the statistical ensemble ${\cal E}_{{\rm p}%
}\left[ {\cal S}_{{\rm st}}\right] $ with nondeterministic elements ${\cal S}%
_{{\rm st}}$, (i.e. the transition ${\cal S}_{{\rm st}}\to {\cal E}_{{\rm p}}%
\left[ {\cal S}_{{\rm st}}\right] $). As far as ${\cal E}_{{\rm p}}\left[ 
{\cal S}_{{\rm st}}\right] $ is a deterministic dynamic system, a
construction of the statistical ensemble ${\cal E}_{{\rm gen}}\left[ {\cal E}%
_{{\rm p}}\right] $, whose elements are the statistical ensembles ${\cal E}_{%
{\rm p}}\left[ {\cal S}_{{\rm st}}\right] $ (i.e. the transition ${\cal E}_{%
{\rm p}}\left[ {\cal S}_{{\rm st}}\right] \to {\cal E}_{{\rm gen}}\left[ 
{\cal E}_{{\rm p}}\right] $), is a comparatively simple formal procedure.
Thus, only the transition ${\cal S}_{{\rm st}}\to {\cal E}_{{\rm p}}\left[ 
{\cal S}_{{\rm st}}\right] $ is conceptual. The most attention will be
concentrated on this procedure.

The state $F\left( t,{\bf x},{\bf p}\right) $ of an ensemble of general form 
${\cal E}_{{\rm gen}}\left[ {\cal S}_{{\rm d}}\right] $ evolves according to
the Liouville equation 
\begin{equation}
{\cal E}_{{\rm gen}}\left[ {\cal S}_{{\rm d}}\right] :\qquad \frac{\partial F%
}{\partial t}+\frac \partial {\partial x^\alpha }\left( \frac{\partial H}{%
\partial p_\alpha }F\right) -\frac \partial {\partial p_\alpha }\left( \frac{%
\partial H}{\partial x^\alpha }F\right) =0  \label{d3.1}
\end{equation}
where $H=H\left( t,{\bf x},{\bf p}\right) $ is the Hamilton function for the
dynamic system ${\cal S}_{{\rm d}}$. A summation is made over repeated Greek
indices from $1$ to $n$.

Dynamic equations for the statistical ensemble of a special form ${\cal E}_{%
{\rm p}}\left[ {\cal S}_{{\rm d}}\right] $ have the form 
\begin{equation}
\frac {\partial\rho } {\partial t}+\frac \partial {\partial x^\alpha }\left[
\rho \frac{\partial H}{\partial p_\alpha }\left( t,{\bf x},{\bf p}\right) %
\right] _{{\bf p=P}}=0  \label{d3.2a}
\end{equation}
\begin{equation}
\frac \partial {\partial t}\left( \rho P_\beta \right) +\frac \partial
{\partial x^\alpha }\left( \rho P_\beta \left[ \frac{\partial H}{\partial
p_\alpha }\left( t,{\bf x},{\bf p}\right) \right] _{{\bf p=P}}\right) +\rho 
\frac{\partial H}{\partial x^\beta }\left( t,{\bf x},{\bf P}\right)
=0,\qquad \beta =1,2,...n  \label{d3.2b}
\end{equation}
Let us interpret $\rho $ as a particle density, and ${\bf v}=\partial
H/\partial {\bf p}$ as a generalized velocity. Then the equation (\ref{d3.2a}%
) is regarded as a continuity equation, and equations (\ref{d3.2b}) may be
interpret as generalized Euler equations for some fluid without pressure.

The system of equations (\ref{d3.2a}), (\ref{d3.2b}) is closed, but it is
not complete, and it cannot be obtained from the variational principle. Let
us add to the generalized Euler equations the equations 
\begin{equation}
\frac{dx^\beta }{dt}=\frac{\partial H}{\partial P_\beta }\left( t,{\bf x},%
{\bf P}\right) ,\qquad \beta =1,2,...n  \label{d3.3c}
\end{equation}
describing a particle motion in a given velocity field ${\bf v=}\partial
H/\partial {\bf P}$. These equations can be rewritten in the form, known in
hydrodynamics as Lin constraints \cite{L63} 
\begin{equation}
\frac{\partial \xi _\beta }{\partial t}+\frac{\partial H}{\partial P_\alpha }%
\left( t,{\bf x},{\bf P}\right) \partial _\alpha \xi _\beta =0,\qquad \beta
=1,2,...n,\qquad \partial _k\equiv \frac \partial {\partial x^k},\qquad
k=0,1,...n  \label{d3.4c}
\end{equation}
Here $\bxi \left( t,{\bf x}\right) {\bf =}\left\{ \xi _\alpha \left( t,{\bf x%
}\right) \right\} ,\;\;\;\alpha =1,2,...n$ are $n$ independent integrals of
equations (\ref{d3.3c}).

The system of $2n+1$ equations (\ref{d3.2a}), (\ref{d3.2b}), (\ref{d3.4c})
forms a complete system of dynamic equations, describing evolution of the
pure statistical ensemble ${\cal E}_{{\rm p}}\left[ {\cal S}_{{\rm d}}\right]
$. It can be integrated and reduced to a system of $n+2$ equations for $n+2$
variables $\rho ,\varphi ,\bxi $ 
\begin{equation}
b_0[\partial _0\varphi +g^\alpha (\bxi )\partial _0\xi _\alpha ]+H\left( 
{\bf x,P}\right) =0  \label{d3.6c}
\end{equation}
\begin{equation}
\partial _0\rho +\partial _\alpha \left( \rho \frac{\partial H}{\partial
P_\alpha }\left( t,{\bf x},{\bf P}\right) \right) =0  \label{d3.2aa}
\end{equation}
\begin{equation}
\frac{\partial \xi _\beta }{\partial t}+\frac{\partial H}{\partial P_\alpha }%
\left( t,{\bf x},{\bf P}\right) \partial _\alpha \xi _\beta =0,\qquad \beta
=1,2,...n  \label{d3.4cc}
\end{equation}
where $\varphi $ is a new variable, and ${\bf P}$ is expressed via $n$
arbitrary functions ${\bf g}\left( \bxi \right) = \left\{ g^\alpha \left( %
\bxi \right) \right\} ,\;\;\;\alpha =1,2,...n$ of argument $\bxi$. 
\begin{equation}
P_\beta =b_0\left( \partial _\beta \varphi +g^\alpha \left( \bxi \right)
\right) \partial _\beta \xi _\alpha ,\qquad \beta =1,2,...n  \label{d3.5c}
\end{equation}
The validity of the statement on integration of the system (\ref{d3.2a}) (%
\ref{d3.2b}), (\ref{d3.4c}) can be verified either by means of a direct
substitution of (\ref{d3.5c}) into (\ref{d3.2b}), or by use of technique,
developed in \cite{R99}. $b_0$ is an arbitrary constant, which may be
incorporated in the variable $\varphi $ and arbitrary functions ${\bf g}%
\left( \bxi \right) $.

The system of $n+2$ equations (\ref{d3.6c}), (\ref{d3.2aa}), (\ref{d3.4cc})
is complete. It is remarkable in the relation, that it can be described in
terms of many-component complex function $\psi $ (wave function). This
transformation can be carried out, using the Hamilton variational principle.

One can show, that dynamic equations (\ref{d3.6c}), (\ref{d3.2aa}), (\ref
{d3.4cc}) for the pure statistical ensemble ${\cal E}_{{\rm p}}\left[ {\cal S%
}_{{\rm d}}\right] $ of deterministic dynamic systems ${\cal S}_{{\rm d}}$
are derived from the variational principle with the action 
\begin{equation}
{\cal E}_{{\rm p}}\left[ {\cal S}_{{\rm d}}\right] :\qquad {\cal A}[\rho
,\varphi ,\bxi ]=-\int \rho \{H\left( t,{\bf x,p}\right) +p_0\}d^{n+1}x,
\label{d3.4}
\end{equation}
\begin{equation}
p_k =b_0\left[ \partial _k\varphi +g^\alpha (\bxi )\partial _k \xi _\alpha %
\right] ,\qquad \partial _k\equiv \frac\partial {\partial x^k} , \qquad k
=0,1,...,n  \label{d3.5}
\end{equation}
where $\rho ,\varphi ,\bxi $ are dependent variables, considered to be
functions of argument $x=\left\{ x^0,{\bf x}\right\} =\left\{ t,{\bf x}%
\right\} $. $H\left( t,{\bf x,p}\right) $ is the Hamilton function for $%
{\cal S}_{{\rm d}}$. $b_0$ is an arbitrary constant, and $g^\alpha (\bxi %
),\;\;\alpha =1,2,...,n$ are arbitrary functions of argument $\bxi $.
Dynamic variables $\varphi ,\bxi $ are hydrodynamic potentials (Clebsch
potentials). Clebsch \cite{C57,C59} had introduced them for description of
incompressible fluid. The variables $\varphi ,\bxi $ are referred to as
potentials, because the momentum ${\bf p}={\bf P}\left( t,{\bf x}\right) $
is expressed via derivatives of the potentials $\varphi ,\bxi $, as one can
see this from relations (\ref{d3.5}). The Hamilton function $H\left( t,{\bf %
x,p}\right) $ is a function, which determines the form of the action (\ref
{d3.4}), and the variational principle, based on (\ref{d3.4}), may be
referred to as the Hamilton variational principle.

Let us introduce a $k$-component complex function $\psi =\{\psi _\alpha
\},\;\;\alpha =1,2,\ldots k$, defining it by the relations 
\begin{equation}
\psi _\alpha =\sqrt{\rho }e^{i\varphi }u_\alpha (\bxi ),\qquad \psi _\alpha
^{*}=\sqrt{\rho }e^{-i\varphi }u_\alpha ^{*}(\bxi ),\qquad \alpha
=1,2,\ldots k  \label{s5.4}
\end{equation}
\[
\psi ^{*}\psi \equiv \sum_{\alpha =1}^k\psi _\alpha ^{*}\psi _\alpha 
\]
where (*) means a complex conjugate, $u_\alpha (\bxi )$, $\;\alpha
=1,2,\ldots k$ are functions of only variables $\bxi $. They satisfy the
relations 
\begin{equation}
-\frac i2\sum_{\alpha =1}^k(u_\alpha ^{*}\frac{\partial u_\alpha }{\partial
\xi _\beta }-\frac{\partial u_\alpha ^{*}}{\partial \xi _\beta }u_\alpha
)=g^\beta (\bxi ),\qquad \beta =1,2,...n,\qquad \sum_{\alpha =1}^ku_\alpha
^{*}u_\alpha =1  \label{s5.5}
\end{equation}
$k$ is such a natural number that equations (\ref{s5.5}) admit a solution.
In general, $k$ depends on arbitrary integration functions ${\bf g}%
=\{g^\beta (\bxi )\}$,\ $\beta =1,2,...n.$

It is easy to verify that 
\begin{equation}
\rho =\psi ^{*}\psi ,\qquad p_l(\varphi ,\xi )=-\frac{ib_0}{2\psi ^{*}\psi }%
(\psi ^{*}\partial _l\psi -\partial _l\psi ^{*}\cdot \psi ),\qquad l=0,1,...n
\label{s5.6}
\end{equation}
The variational problem with the action (\ref{d3.4}) appears to be
equivalent to the variational problem with the action functional 
\[
A[\psi ,\psi ^{*}]=\int \left\{\frac{ib_0}2(\psi ^{*}\partial _0\psi
-\partial _0\psi ^{*}\cdot \psi )\right. 
\]
\begin{equation}
\left. -H\left( x,-\frac{ib_0}{2\psi ^{*}\psi }(\psi ^{*}{\bf \nabla }\psi -%
{\bf \nabla }\psi ^{*}\cdot \psi )\right) \psi ^{*}\psi \right\} {\rm d}%
^{n+1}x  \label{s5.8}
\end{equation}
where ${\bf \nabla }=\left\{ \nabla _\alpha \right\} =\left\{ \partial
_\alpha \right\} ,\;\;\alpha =1,2,...n$.

Let us note, that the function $\psi $, considered to be a function of
independent variables $x=\{t,{\bf x}\}$ is very indefinite in the sense,
that the same state $\left\{ \rho \left( t,{\bf x}\right) ,{\bf P}\left( t,%
{\bf x}\right) \right\} $ of the statistical ensemble ${\cal E}_{{\rm p}}%
\left[ {\cal S}_{{\rm d}}\right] $ can be described by various $\psi $%
-functions. There are two reasons for such an indefiniteness. First, the
functions $u_\alpha ({\bf \xi })$ are not determined single-valuedly by the
equations (\ref{s5.5}). Second, their arguments $\bxi $ as functions of $x$
are determined within the relabelling transformation 
\begin{equation}
\xi _\alpha \to \tilde \xi _\alpha =\tilde \xi _\alpha (\bxi ),\qquad \det
\parallel \partial \tilde \xi _\alpha /\partial \xi _\beta \parallel
=1,\qquad \alpha ,\beta =1,2,...n  \label{d1.16}
\end{equation}

Description of the statistical ensemble ${\cal E}_{{\rm p}}\left[ {\cal S}_{%
{\rm d}}\right] $ in terms of the function $\psi $ is more indefinite, than
a description in terms of hydrodynamic potentials $\bxi $. Information on
initial and boundary conditions, contained in functions ${\bf g}(\bxi )$, is
lost at the description in terms of $\psi $-function.

The dynamic equations have the form 
\begin{equation}
\delta \psi _\beta ^{*}:\qquad \left[ ib_0\partial _0-H+\frac{\partial H}{%
\partial p_\alpha }p_\alpha +\frac{ib_0}2\left( \frac{\partial H}{\partial
p_\alpha }\nabla _\alpha +\nabla _\alpha \frac{\partial H}{\partial p_\alpha 
}\right) \right] \psi _\beta =0,\qquad \beta =1,2,...k  \label{s5.7}
\end{equation}
\begin{equation}
\delta \psi _\beta :\qquad \left[ -ib_0\partial _0-H+\frac{\partial H}{%
\partial p_\alpha }p_\alpha -\frac{ib_0}2\left( \frac{\partial H}{\partial
p_\alpha }\nabla _\alpha +\nabla _\alpha \frac{\partial H}{\partial p_\alpha 
}\right) \right] \psi _\beta ^{*}=0,\qquad \beta =1,2,...k  \label{s5.7a}
\end{equation}
where $H=H\left( x,{\bf p}\right) $ and $\frac{\partial H}{\partial p_\alpha 
}\left( x,{\bf p}\right)$ are considered to be multiplication operators by
these quantities, the expression (\ref{s5.6}) has to substituted instead of $%
{\bf p}$, and thereafter the operator ${\bf \nabla }$ has to act. In
general, dynamic equations (\ref{s5.7}), (\ref{s5.7a}) are nonlinear with
respect to $\psi $-function, although they appear to be linear in some
cases. In these cases the dynamic equations can be solved easily.

The number $k$ of the $\psi $-function components in the action (\ref{s5.8})
is arbitrary. A formal variation of the action with respect to $\psi _\alpha 
$ and $\psi _\alpha ^{*},\quad \alpha =1,2,\ldots k$ leads to $2k$ real
dynamic equations, but not all of them are independent. There are such
combinations of variations $\delta \psi _\alpha $, $\delta \psi _\alpha ^{*}$%
, $\alpha =1,2,\ldots k$, do not change expressions (\ref{s5.6}). Such
combinations of variations $\delta \psi _\alpha $, $\delta \psi _\alpha ^{*}$%
, $\alpha =1,2,\ldots k$ do not change the action (\ref{s5.8}), and
corresponding combinations of dynamic equations $\delta {\cal A}/\delta \psi
_\alpha =0$, $\delta {\cal A}/\delta \psi _\alpha ^{*}=0$ are identities. It
associates with a connection between dynamic equations.

Thus, the number of equations increases at increase of the number $k$, but
the number of independent dynamic equations remains the same. The number $k$
is restricted from below by the constraint, that the equations (\ref{s5.5})
have a solution. in other words, the minimal number $k_m$ of the $\psi $%
-function components depends on the form of functions ${\bf g}(\bxi )$, i.e.
on the initial conditions. This number $k_m$ associates with a kinematic
spin ( $k$-spin) $s=2k_m+1$ of the ensemble state \cite{R99}.

$\psi $-function and $k$-spin remind respectively wave function and spin. $%
\psi $-function coincides with the wave function, provided dynamic equations
(\ref{s5.7}), (\ref{s5.7a}) becomes linear. It appears to be possible for a
pure statistical ensemble ${\cal E}_{{\rm p}}\left[ {\cal S}_{{\rm st}}%
\right] $ of stochastic systems ${\cal S}_{{\rm st}}$. In this case the $k$%
-spin associates with the spin of a particle, but the $k$-spin remains to be
a property of the statistical ensemble ${\cal E}_{{\rm p}}\left[ {\cal S}_{%
{\rm st}}\right] $ (i.e. a collective property), whereas in quantum
mechanics the spin is considered to be a property of a single particle.

For this reason one should note that in quantum mechanics the spin is a
property of a single particle not always. In the paper \cite{R95} the
properties of two dynamic systems ${\cal S}_{{\rm S}}$ and ${\cal S}_{{\rm P}%
}$, described respectively by the Schr\"odinger equation and by the Pauli
one, were analyzed. It appears that in the classical approximation both
dynamic systems can be interpreted as pure statistical ensembles
respectively ${\cal E}_{{\rm S}}\left[ {\cal S}_{{\rm d}}\right] $ and $%
{\cal E}_{{\rm P}}\left[ {\cal S}_{{\rm d}}\right] $, whose elements appear
to be the same dynamic system ${\cal S}_{{\rm d}}$. The statistical
ensembles ${\cal E}_{{\rm S}}\left[ {\cal S}_{{\rm d}}\right] $, ${\cal E}_{%
{\rm P}}\left[ {\cal S}_{{\rm d}}\right] $ differ only in their structure,
i.e. in a choice of functions ${\bf g}(\bxi )$.

Thus, analysis of the description methods of the pure statistical ensemble
description shows that the wave function and spin are not specific quantum
objects. The wave function is simply a set of complex potentials, and it
contains not more mysticism, than electromagnetic potentials.

\section{Pure statistical ensemble of stochastic systems}

Let us consider a statistical ensemble$_{{\rm p}}\left[ {\cal S}_{{\rm st}}%
\right] $ of stochastic systems ${\cal S}_{{\rm st}}$. There are no dynamic
equations for ${\cal S}_{{\rm st}}$, and dynamic equations for ${\cal E}_{%
{\rm p}}\left[ {\cal S}_{{\rm st}}\right] $ cannot be derived from dynamic
equations for ${\cal S}_{{\rm st}}$. But we believe that dynamic equations
for ${\cal E}_{{\rm p}}\left[ {\cal S}_{{\rm st}}\right] $ do exist, as far
as experiments with statistical ensembles of stochastic particles ${\cal S}_{%
{\rm st}}$ are reproducible.

Let us consider a motion of stochastic particle ${\cal S}_{{\rm st}}$ as a
result of interaction between a deterministic particle ${\cal S}_{{\rm d}}$
and some stochastic agent, which perturbs motion of ${\cal S}_{{\rm d}}$ and
make it to be stochastic. To derive dynamic equations for ${\cal E}_{{\rm p}}%
\left[ {\cal S}_{{\rm st}}\right] $, some suppositions on properties of this
agent are to be made, because it is impossible to derive dynamic equations
for ${\cal E}_{{\rm p}}\left[ {\cal S}_{{\rm st}}\right] $ from nothing. If $%
{\cal S}_{{\rm st}}$ is a Brownian particle, moving in a gas, one supposes
that the Brownian particle collides with gas molecules, and these collisions
make the Brownian particle motion to be stochastic. These collisions are
supposed to be independent and random. The Brownian particle motion appears
to be a Markovian process. The dynamic system ${\cal E}_{{\rm p}}\left[ 
{\cal S}_{{\rm st}}\right] $ appears to be dissipative, and there is no
variational principle for it.

Such a way of description is not suit for description of the geometric
stochasticity influence, because the random component of the particle motion
is relativistic, the probabilistic statistical description cannot be used.
It is supposed that the stochastic agent influence manifests in the
averaging of parameters of the Hamilton function $H$, describing motion of $%
{\cal S}_{{\rm d}}$. These parameters start to depend on the state of the
statistical ensemble ${\cal E}_{{\rm p}}\left[ {\cal S}_{{\rm d}}\right] $,
i.e. on the collective variable $\rho$. Elements of the statistical ensemble
start to interact between themselves and stop to be independent. The dynamic
system ${\cal E}_{{\rm p}}\left[ {\cal S}_{{\rm d}}\right] $ stops to be a
statistical ensemble and turns to a dynamic system ${\cal E}_{{\rm red}}%
\left[ {\cal S}_{{\rm d}}\right] $, which will be referred to as reduced
ensemble (the word "statistical" is omitted).

For a free relativistic deterministic particle the Hamilton function has the
form 
\begin{equation}
H\left( x,{\bf p}\right) =\sqrt{m^{2}c^{4}+{\bf p}^{2}c^{2}}  \label{e4.1}
\end{equation}
where the mass $m$ is the only parameter of Hamiltonian of the system ${\cal %
S}_{{\rm d}}$. The variational principle (\ref{d3.4}) for dynamic system $%
{\cal E}_{{\rm p}}\left[ {\cal S}_{{\rm d}}\right] $ has the form 
\begin{equation}
{\cal A}[\rho ,\varphi ,\bxi]=\int \rho \{-\sqrt{m^{2}c^{4}+{\bf p}^{2}c^{2}}%
-b_{0}[\partial _{0}\varphi +g^{\alpha }(\bxi)\partial _{0}\xi _{\alpha }]\}%
{\rm d}^{4}x,  \label{e4.2}
\end{equation}
where ${\bf p}$ is given by the relation (\ref{d3.5}) with $n=3$. After
averaging \cite{R91,R1995}, which is produced with taking into account the
world function (\ref{a1.2}), (\ref{a1.3}), the effective mass $m$ of the
particle changes 
\begin{equation}
m^{2}\rightarrow m_{{\rm q}}^{2}=m^{2}+\frac{\hbar ^{2}}{4c^{2}}\left( {\bf %
\nabla }\ln \rho \right) ^{2}  \label{e4.5}
\end{equation}
After substitution $m^{2}\rightarrow m_{{\rm q}}^{2}$ the action takes the
form 
\begin{equation}
{\cal A}[\rho ,\varphi ,\bxi]=\int \rho \{-\sqrt{m^{2}c^{4}+{\bf p}^{2}c^{2}+%
\frac{\hbar ^{2}c^{2}}{4}\left( {\bf \nabla }\ln \rho \right) ^{2}}%
-b_{0}[\partial _{0}\varphi +g^{\alpha }(\bxi)\partial _{0}\xi _{\alpha }]\}%
{\rm d}^{4}x,  \label{e4.4}
\end{equation}
The Hamilton function 
\begin{equation}
H_{{\rm eff}}\left( x,{\bf p}\right) =\sqrt{m^{2}c^{4}+\frac{\hbar ^{2}c^{2}%
}{4}\left( {\bf \nabla }\ln \rho \right) ^{2}+{\bf p}^{2}c^{2}}  \label{e4.6}
\end{equation}
appears to be invariant with respect to transformation $\rho \rightarrow
a\rho $, $a=$const.

The action (\ref{e4.4}) is an action for some statistical ensemble, because
for the action (\ref{e4.4}) the condition (\ref{a3.1}) of independence on
the number of elements takes the form 
\begin{equation}
{\cal A}\left[ a\rho ,\varphi ,{\bf \xi }\right] =a{\cal A}\left[ \rho
,\varphi ,{\bf \xi }\right] ,\qquad a=\mbox{const},\qquad a>0.  \label{e4.7}
\end{equation}
This condition is satisfied, but now the action (\ref{e4.4}) cannot be
interpreted as an action for a pure statistical ensemble, whose elements are
some deterministic systems ${\cal S}_{{\rm d}}$, because these dynamic
systems ${\cal S}_{{\rm d}}$ interact between themselves and are not
independent. It means that the action (\ref{e4.4}) can be and must be
interpreted as an action for a pure statistical ensemble ${\cal E}_{{\rm p}}%
\left[ {\cal S}_{{\rm st}}\right] $, whose elements are stochastic systems $%
{\cal S}_{{\rm st}}$.

In the nonrelativistic approximation the action (\ref{e4.4}) has the form 
\begin{equation}
{\cal A}[\rho ,\varphi ,\bxi]=\int \rho \{-mc^{2}-\frac{{\bf p}^{2}}{2m}-%
\frac{\hbar ^{2}}{8m}\left( {\bf \nabla }\ln \rho \right)
^{2}-b_{0}[\partial _{0}\varphi +g^{\alpha }(\bxi)\partial _{0}\xi _{\alpha
}]\}{\rm d}^{4}x,  \label{e4.11}
\end{equation}
where ${\bf p}$ is determined by the relation (\ref{d3.5}). The action (\ref
{e4.11}) cannot be interpreted as an action for a statistical ensemble $%
{\cal E}_{{\rm p}}\left[ {\cal S}_{{\rm d}}\right] $ of deterministic
systems ${\cal S}_{{\rm d}}$, but it can be regarded as an action for the
set ${\cal E}_{{\rm red}}\left[ {\cal S}_{{\rm d}}\right] $ of deterministic
systems ${\cal S}_{{\rm d}}$, interacting between themselves by means of the
potential energy 
\begin{equation}
E_{{\rm pot}}=\frac{{\bf p}_{{\rm st}}^{2}}{2m}=\frac{\hbar ^{2}}{8m}\left( 
{\bf \nabla }\ln \rho \right) ^{2}.  \label{e4.11a}
\end{equation}
where ${\bf p}_{{\rm st}}=-\hbar {\bf \nabla }\ln \rho /2$ is the mean
momentum of the stochastic component of the particle motion. Thus, on the
one hand, (\ref{e4.11}) is an action for the statistical ensemble ${\cal E}_{%
{\rm p}}\left[ {\cal S}_{{\rm st}}\right] $ of stochastic systems ${\cal S}_{%
{\rm st}}$, but on the other hand, (\ref{e4.11}) is an action for the set $%
{\cal E}_{{\rm red}}\left[ {\cal S}_{{\rm d}}\right] $ of interacting
deterministic systems ${\cal S}_{{\rm d}}$. It means that one can set up a
correspondence between the stochasticity character and the form of
deterministic systems ${\cal S}_{{\rm d}}$ interaction. Then one can label
the stochasticity character by the form of this interaction. Essentially,
such a reduction of a stochasticity to an interaction is the only possible
way of a mathematical description of a stochasticity. It is described by the
relation 
\begin{equation}
{\cal E}_{{\rm p}}\left[ {\cal S}_{{\rm st}}\right] ={\cal E}_{{\rm red}}%
\left[ {\cal S}_{{\rm d}}\right]   \label{e4.11b}
\end{equation}

In terms of $\psi $-function (\ref{s5.4}) the action (\ref{e4.11}) is
written in the form 
\begin{eqnarray}
A[\psi ,\psi ^{\ast }] &=&\int \{\frac{ib_{0}}{2}(\psi ^{\ast }\partial
_{0}\psi -\partial _{0}\psi ^{\ast }\cdot \psi )-mc^{2}\rho -\frac{\hbar
^{2}\left( {\bf \nabla }\rho \right) ^{2}}{8m\rho }  \nonumber \\
&&+\frac{b_{0}^{2}}{8\rho m}(\psi ^{\ast }{\bf \nabla }\psi -{\bf \nabla }%
\psi ^{\ast }\cdot \psi )^{2}\}{\rm d}^{4}x,  \label{e4.12}
\end{eqnarray}
where $\rho \equiv \psi ^{\ast }\psi $.

Let the function $\psi $ have $k$ components. Regrouping components of the
function $\psi $ of the action (\ref{e4.12}), one obtains it in the form 
\[
{\cal A}[\psi ,\psi ^{*}]=\int \{\frac{ib_0}2(\psi ^{*}\partial _0\psi
-\partial _0\psi ^{*}\cdot \psi )-\frac{b_0^2}{2m}{\bf \nabla }\psi
^{*}\cdot {\bf \nabla }\psi 
\]
\begin{equation}
+\frac{b_0^2}4\sum\limits_{\alpha ,\beta =1}^kQ_{\alpha \beta ,\gamma
}^{*}Q_{\alpha \beta ,\gamma }\rho +\frac{b_0^2-\hbar }{8\rho m}^2(\nabla
\rho )^2-mc^2\rho \}{\rm d}^4x,\qquad \rho \equiv \psi ^{*}\psi  \label{s5.9}
\end{equation}
where a summation over $\gamma$ is supposed from $1$ to $3$, 
\begin{equation}
Q_{\alpha \beta ,\gamma }={\frac 1{\psi ^{*}\psi }}\left| 
\begin{array}{cc}
\psi _\alpha & \psi _\beta \\ 
\partial _\gamma \psi _\alpha & \partial _\gamma \psi _\beta
\end{array}
\right| ,\qquad \alpha ,\beta =1,2,\ldots k\qquad \gamma =1,2,3
\label{s5.11}
\end{equation}
and $Q_{\alpha \beta ,\gamma }^{*}$ is the complex conjugate to the quantity 
$Q_{\alpha \beta ,\gamma }$.

In the simplest case, when the $\psi $-function has only one component, all
quantities $Q_{11,\gamma }={0},\quad \gamma =1,2,3$, and the ensemble
particle motion is irrotational. Then the action (\ref{s5.9}) reduces to the
form 
\[
{\cal A}[\psi ,\psi ^{*}]=\int \{\frac{ib_0}2(\psi ^{*}\partial _0\psi
-\partial _0\psi ^{*}\cdot \psi )-\frac{b_0^2}{2m}{\bf \nabla }\psi
^{*}\cdot {\bf \nabla }\psi 
\]
\begin{equation}
-mc^2\rho +\frac{b_0^2-\hbar ^2}{8\rho m}(\nabla \rho )^2\}{\rm d}^4x,\qquad
\rho \equiv \psi ^{*}\psi  \label{e4.13}
\end{equation}

Due to the last term in the action (\ref{e4.13}) the dynamic equation,
generated by the action (\ref{e4.13}) is nonlinear, except for the case,
when $b_0^2=\hbar ^2 $, although $b_0$ is an integration constant, and the
action (\ref{e4.13}) describes the same dynamic system for any value of $b_0$%
. Equating the arbitrary constant $b_0$ to $\hbar $, $\left( b_0=\hbar
\right) $, one obtains instead of (\ref{e4.13}) 
\begin{equation}
{\cal A}[\psi ,\psi ^{*}]=\int \{\frac{i\hbar }2(\psi ^{*}\partial _0\psi
-\partial _0\psi ^{*}\cdot \psi )-\frac{\hbar ^2}{2m}{\bf \nabla }\psi
^{*}\cdot {\bf \nabla }\psi -mc^2\psi ^{*}\cdot \psi \}{\rm d}^4x
\label{e4.14}
\end{equation}

It is easy to see that the dynamic equation, generated by the action (\ref
{e4.14}) is linear. After the substitution $\psi\to \exp{(-imc^2t/\hbar )}%
\psi$, removing the rest mass, the equation turns to the Schr\"odinger
equation in its conventional form 
\begin{equation}
i\hbar \partial _0\psi +\frac{\hbar ^2}{2m}{\bf \nabla }^2\psi =0.
\label{е4.15}
\end{equation}

The constant $b_0$ describes the phase scale of the $\psi $-function, and
the transformation of the $\psi $-function phase 
\begin{equation}
\psi \to \tilde \psi =|\psi |\exp \left( \frac{\tilde b_0}{b_0}\log \frac
\psi {|\psi |}\right) ,  \label{e4.16}
\end{equation}
changes the constant $b_0$ to the constant $\tilde b_0$ in the action (\ref
{e4.13}). The actions (\ref{e4.13}) and (\ref{e4.14}) distinguish very
strongly between themselves, although both describe the same dynamic system.
The action (\ref{e4.13}) contains only one quantum term, i.e. the term,
containing $\hbar $, and setting $\hbar =0$, one passes automatically from
quantum description to classical one. Vice versa, in the action almost all
terms are quantum, and one cannot set $\hbar =0$, because then any dynamic
system description disappears. For derivation of classical description from
the action (\ref{e4.14}) it is to use subtle methods of quasi-classical
description. Linearity of dynamic equation, arising at the transition from
the action (\ref{e4.13}) to the action (\ref{e4.14}), looks rather as a
happy chance, than a manifestation of quantum-mechanical principle of
dynamic equation linearity.

Describing stochastic systems ${\cal S}_{{\rm st}}$ by means of the action (%
\ref{e4.14}), one can interpret the quantity $\psi ^{*}({\bf x})\psi ({\bf x}%
)$ as the probability density to discover a particle at the point ${\bf x}$.
It is connected with the fact that the quantity $\psi ^{*}({\bf x})\psi (%
{\bf x})$ is non-negative, and integral from it is conserved due to dynamic
equation. The probability density, introduced in such a way, is very
convenient, but it has not a direct relation to the statistical description.
In general, consideration of the action (\ref{e4.14}) for the dynamic system 
${\cal E}_{{\rm red}}\left[ {\cal S}_{{\rm d}}\right] $ does not associate
with conventional conception of the statistical description.

One can show \cite{R98}, that setting of a dynamic system, i.e. setting of
the action (\ref{e4.14}), is enough for a description of all quantum effects
(diffraction, interference, tunneling, uncertainty relation, determination
of eigenvalues for stationary states, etc.). In other words, if the dynamic
system (\ref{e4.14}) is given, one can describe all quantum effects without
a reference to quantum mechanics principles. This statement is valid not
only in the special case of the action (\ref{e4.14}), but in the general
case of the action, appeared as a corollary of statistical description. This
statement finishes the logical scheme of the research program Copernicus-2.

Thus, in the non-relativistic approximation the program Copernicus-2 gives
the quantum mechanical description, basing only on the space-time geometry
without QM principles. The general relativistic case has been developed
insufficiently, but the statistical description, which leads to the dynamic
system ${\cal S}_{{\rm KG}}$, described by the Klein-Gordon equation has
been obtained in \cite{R98}. For its derivation one needs to use a
relativistic version, where nonrelativistic effective mass $m_{{\bf q}}$,
given by the relation (\ref{e4.5}) is substituted by its relativistic
version, and the temporal component $j^0=\rho $ is substituted by
corresponding relativistic invariant $j^0/H$. But the nonrelativistic
Hamilton variational principle is slightly suit for dealing with
relativistic quantities. It is more convenient to use the Lagrange
variational principle equivalent to (\ref{d3.4}) 
\begin{equation}
{\cal E}\left[ {\cal S}_{{\rm d}}\right] :\qquad {\cal A}[j,\varphi ,\bxi %
]=\int \{L(x,\frac{{\bf j}}{j^0})j^0-b_0j^i[\partial _i\varphi +g^\alpha (%
\bxi )\partial _i\xi _\alpha ]\}{\rm d}^{n+1}x,  \label{e4.17}
\end{equation}
where $L(x,\frac{d{\bf x}}{dt})$ is the Lagrangian of the system ${\cal S}_{%
{\rm d}}$ and $\{j^0,{\bf j}\}=\left\{ j^i\right\} ,\;\;i=0,1,...n$ is the
flux of particle ${\cal S}_{{\rm d}}$ in the statistical ensemble ${\cal E}%
\left[ {\cal S}_{{\rm d}}\right] $. Then the variational principle for the
statistical ensemble ${\cal E}\left[ {\cal S}_{{\rm st}}\right] ={\cal E}_{%
{\rm red}}\left[ {\cal S}_{{\rm d}}\right] $ of stochastic systems ${\cal S}%
_{{\rm st}}$ takes the form 
\begin{equation}
{\cal E}\left[ {\cal S}_{{\rm st}}\right] :\qquad {\cal A}[j,\varphi ,\bxi %
,\kappa ]=\int \{-mcK\sqrt{j^ig_{ik}j^k}-b_0j^i[\partial _i\varphi +g^\alpha
(\bxi )\partial _i\xi _\alpha ]\}{\rm d}^3x,  \label{e4.18}
\end{equation}
\begin{equation}
m_{{\rm q}}=mK,\qquad K\equiv \sqrt{1+\lambda ^2(\partial _l\kappa ^l+\kappa
^l\kappa _l)},\qquad \partial _k\equiv \partial /\partial x^k,  \label{e4.19}
\end{equation}
where $g_{ik}=$diag$\{c^2,-1,-1,-1\}$ is the metric tensor, $m$ is the
particle mass and $\lambda \equiv \hbar /mc$ is its Compton wavelength. $%
\bxi =\{\xi _\alpha \},\quad \alpha =1,2,3$; and $\kappa ^l=\kappa
^l(x),\quad x=\{x^l\},\quad l=0,1,2,3$. A summation is made over repeating
indices, for Latin indices from $0$ to $3$, and for Greek ones from $1$ to $%
3 $. Here the effective mass $m_{{\rm q}}=mK$ is expressed via some $\kappa $%
-field, describing interaction of particles ${\cal S}_{{\rm d}}$ in the
dynamic system ${\cal E}_{{\rm red}}\left[ {\cal S}_{{\rm d}}\right] $. At
the same time the $\kappa $-field describes stochasticity of the systems $%
{\cal S}_{{\rm st}}$.

It follows \cite{R98} from dynamic equations, that the $\kappa $-field has a
potential, designed by means of $\frac 12\ln \rho $, i.e. $\kappa _l=\frac
12\partial _l\ln \rho .$ Then one can introduce the $\psi $-function by
means of relations (\ref{s5.4}). In the simplest case, when $\psi $-function
has only one component, the dynamic equation for it coincides with the
Klein-Gordon equation and with the Schr\"odinger one in the nonrelativistic
approximation.

The $\kappa $-field has all characteristic properties of a field, i.e. it
has a proper energy, it can exist in the absence of a matter, i.e. at $%
j^k=0,\quad k=0,1,2,3$. Besides, it enables to produce pairs
particle--antiparticle and is responsible for quantum effects. It means,
that at $\kappa ^i\equiv 0,\quad i=0,1,2,3$ the statistical ensemble ${\cal E%
}\left[ {\cal S}_{{\rm st}}\right] $ turns to the statistical ensemble $%
{\cal E}\left[ {\cal S}_{{\rm d}}\right] $.

\section{Concluding remarks}

The research program Copernicus-2 is more perfect logically, than Ptolemy-2,
because it was founded on the basis of more general and perfect conceptions
of geometry and statistical description. It is important to understand, that
these more general conceptions are not a result of some successful
hypotheses, or restrictions. On the contrary, the larger generality and
efficiency of the new conceptions of geometry and statistical description
appear as a corollary of a removal of unfounded constraints, used earlier.
The new conception of geometry does not use the concept of a curve, because
it is too restrictive. The new conception of the statistical description
does not use concept of probability and that of probability density, because
they are also too restrictive. It is the point, that {\it simultaneous
application} of both T-geometry and dynamical conception of statistical
description is very important also. A use of only T-geometry explains the
origin of quantum stochasticity, but it does not admit to reconstruct the
mathematical technique of quantum mechanics. A use of only dynamical
conception of statistical descriptions admits one to derive the mathematical
technique of quantum mechanics, but it does not explains the origin of
quantum stochasticity, and does not permit to develop this technique in the
''geometrical direction'', that is characteristic for the whole development
of physics in the last century

From the fact, that the research program Copernicus-2 explains quantum
effects without a reference to additional hypotheses (QM principles), it
follows that Copernicus-2 is more logically consistent, than Ptolemy-2. The
last program uses inadequate space-time geometry, which is to be corrected.
But the program Ptolemy-2 works almost hundred years. All descriptions of
quantum phenomena and corresponding calculations are produced in terms of
quantum mechanics. Vast factual data were collected, and revision of these
data is difficult and undesirable. In this connection it is very important
to know, to what degree a transition from the program Ptolemy-2 to the
program Copernicus-2 concerns existing results, obtained on the basis of
quantum mechanics.

To estimate this, it is useful to turn to an experience of interplay between
the axiomatic thermodynamics and statistical physics, which founded
thermodynamics and determined limits of applicability of its relations. This
experience shows that restrictions imposed by the statistical physics,
concern only a small part of thermodynamics results. Nothing changed in the
field, where thermodynamics was used for practical goals. One should expect
that a transition to the program Copernicus-2 will change nothing in
nonrelativistic quantum mechanics, which has been developed mostly and has
practical applications. In the relativistic quantum mechanics, especially in
the elementary particle theory the changes may be essential.

Let us note an important problem, connected with the dynamic system ${\cal S}%
_{{\rm D}}$, described by the Dirac equation, or by the action 
\begin{equation}
{\cal S}_{{\rm D}}:\qquad {\cal A}_{{\rm D}}[\bar \psi ,\psi ]=\int (-m\bar
\psi \psi +\frac i2\hbar \bar \psi \gamma ^l\partial _l\psi -\frac i2\hbar
\partial _l\bar \psi \gamma ^l\psi )d^4x  \label{d4.1}
\end{equation}
where $\psi $ is a four-component complex wave function. It is known that
the Dirac equation is a relativistic equation, but the dynamic system ${\cal %
S}_{{\rm D}}$ is not relativistic, and it is very unexpected. This fact was
discovered at the analysis of dynamic system ${\cal S}_{{\rm D}}$ \cite{R995}%
, undertaken for investigation of what a geometrical object is associated
with ${\cal S}_{{\rm D}}$. The meaning of Dirac matrices $\gamma ^i$ in the
action (\ref{d4.1}) is obscure. They were eliminated, and the system ${\cal S%
}_{{\rm D}}$ was investigated in tensor variables $j^l,S^l$, $(l=0,1,2,3)$, $%
\varphi ,\kappa $, determined by the relations 
\begin{equation}
j^l=\bar \psi \gamma ^l\psi ,\qquad l=0,1,2,3,\qquad \bar \psi =\psi
^{*}\gamma ^0,  \label{d4.2}
\end{equation}
\begin{equation}
S^l=i\bar \psi \gamma _5\gamma ^l\psi ,\qquad l=0,1,2,3,\qquad \gamma
_5=\gamma ^{0123}\equiv \gamma ^0\gamma ^1\gamma ^2\gamma ^3,  \label{d4.3}
\end{equation}
Here $\gamma ^l$, $l=0,1,2,3$ are Dirac $\gamma $-matrices, satisfying the
commutation relations 
\begin{equation}
\gamma ^i\gamma ^k+\gamma ^k\gamma ^i=2g^{ik},\qquad i,k=0,1,2,3,
\label{d4.5}
\end{equation}
where $g^{ik}$ =diag$(1,-1,-1,-1)$ is the metric tensor. Only two of
components of the pseudovector $S^l$ are independent, because there are two
identities 
\begin{equation}
S^lS_l\equiv -j^lj_l,\qquad j^lS_l\equiv 0.  \label{d4.6}
\end{equation}

To describe ${\cal S}_{{\rm D}}$ in tensor variables, the change of
variables is made 
\begin{equation}
\psi =Ae^{i\varphi +{\frac 12}\gamma _5\kappa }\exp{(-\frac i2\gamma _5{\bf %
\sigma \eta })}\left( {\bf \sigma n}\right) \Pi ,\qquad \bar \psi =A\Pi
\left( {\bf \sigma n}\right) \exp{(-\frac i2\gamma _5{\bf \sigma \eta )}}%
e^{-i\varphi +{\frac 12}\gamma _5\kappa }  \label{d4.7}
\end{equation}
where $\Pi $ is the zero devisor 
\begin{equation}
\Pi =\frac 14\left( 1+\gamma ^0\right) \left( 1+{\bf z\sigma }\right)
,\qquad {\bf z}=\left\{ z^1,z^2,z^3\right\} ,\qquad {\bf z}^2=1  \label{d4.8}
\end{equation}
\begin{equation}
{\bf \sigma }=\left\{ \sigma _1,\sigma _2,\sigma _3,\right\} ,\qquad \sigma
_1=-i\gamma ^2\gamma ^3,\qquad \sigma _2=-i\gamma ^3\gamma ^1,\qquad \sigma
_3=-i\gamma ^1\gamma ^2  \label{d4.9}
\end{equation}
The variables $A,{\bf \eta }=\left\{ \eta ^1,\eta ^2,\eta ^3\right\} ,\;{\bf %
n}=\left\{ n^1,n^2,n^3\right\} ,$ $\left( {\bf n}^2=1\right) $ are six
intermediate variables, and ${\bf z}$ is a constant unite 3-vector.
Substituting (\ref{d4.7}) in (\ref{d4.1}) and using (\ref{d4.2}), (\ref{d4.3}%
), one can express the action (\ref{d4.1}) in terms of tensor variables $%
j^l,S^l,\kappa ,\varphi $ with eight independent real components.

One expected that after transformation to tensor variables $j^l,S^l$, $%
\varphi ,\kappa $, one succeeded to write the action (\ref{d4.1}) in the
relativistically covariant form. But it failed. The action and dynamic
equations are written in the relativistically covariant form only after
introduction of constant unit timelike 4-vector $f^i$. This 4-vector is an
absolute object in the sense of Anderson \cite{A67}. (Note that the constant
vector ${\bf z}$ is another absolute object, but it appears to be
fictitious.) The 4-vector $f^i$ describes separation of the space-time into
space and time. In other words, the dynamic system ${\cal S}_{{\rm D}}$
appears to be nonrelativistic. Of course, it is nonrelativistic at the
description in terms of the wave function $\psi $ also, but in this case the
4-vector $f^i$ is absorbed by other absolute objects ($\gamma $-matrices),
and one cannot discover it at once (see discussion in \cite{R995}). One may
think that appearance of $f^i$ is a result of a calculation mistake
(transformation of the action (\ref{d4.1}) to tensor variables is rather
bulky). But the same timelike vector $f^i$ appears in a more simple case of
two-dimensional space-time, when a transformation of the system ${\cal S}_{%
{\rm D}}$ to the dynamic system ${\cal S}_{{\rm KG}}$, described by the
Klein-Gordon equation, is possible \cite{R999}. Unfortunately, this
circumstance forces one to think that the conclusion on nonrelativistic
character of dynamic system ${\cal S}_{{\rm D}}$ is valid.

Thus, the dynamic system ${\cal S}_{{\rm D}}$ is nonrelativistic, and it is
a serious test for both research programs Ptolemy-2 and Copernicus-2.
Establishing of reasons of this circumstance could advance us in explanation
of microcosm phenomena.

The research programs Ptolemy-2 and Copernicus-2 have guided the different
development of further fundamental investigations, and therein lies the main
difference between them. The key word for further investigation under
program Ptolemy-2 is {\it linearity}, whereas for the program Copernicus-2
the key word is {\it geometrization}. 

\end{document}